\documentclass[fleqn,usenatbib]{mnras}

\usepackage{newtxtext,newtxmath}

\usepackage[T1]{fontenc}

\DeclareRobustCommand{\VAN}[3]{#2}
\let\VANthebibliography\thebibliography
\def\thebibliography{\DeclareRobustCommand{\VAN}[3]{##3}\VANthebibliography}

\usepackage{graphicx}	
\usepackage{amsmath}	
\usepackage{xcolor}
\usepackage[normalem]{ulem}

\definecolor{seagreen}{rgb}{0.18, 0.55, 0.34}

\title[Magnetar-Boosted Kilonova]{Constraining the long-lived supramassive neutron stars by magnetar boosted kilonovae}

\author[Hao Wang et al.]{
Hao Wang,$^{1}$\thanks{E-mail: wang4145@purdue.edu}
Paz Beniamini,$^{2,3,4}$
Dimitrios Giannios$^{1}$\thanks{E-mail: dgiannios@purdue.edu}
\\
$^{1}$Department of Physics and Astronomy, Purdue University, 525 Northwestern Avenue, West Lafayette, IN 47907, USA\\
$^{2}$Department of Natural Sciences, The Open University of Israel, P.O Box 808, Ra'anana 4353701, Israel\\
$^{3}$Astrophysics Research Center of the Open university (ARCO), The Open University of Israel, P.O Box 808, Ra'anana 4353701, Israel\\	
$^{4}$Department of Physics, The George Washington University, 725 21st Street NW, Washington, DC 20052, USA
}

\date{Accepted XXX. Received YYY; in original form ZZZ}

\pubyear{2023}

\begin{document}
\label{firstpage}
\pagerange{\pageref{firstpage}--\pageref{lastpage}}
\maketitle

\begin{abstract}
Kilonovae are optical transients following the merger of neutron star binaries, which are powered by the r-process heating of merger ejecta. However, if a merger remnant is a long-lived supramassive neutron star supported by its uniform rotation, it will inject energy into the ejecta through spindown power. The energy injection can boost the peak luminosity of a kilonova by many orders of magnitudes, thus significantly increasing the detectable volume. Therefore, even if such events are only a small fraction of the kilonovae population, they could dominate the detection rates. However, after many years of optical sky surveys, no such event has been confirmed. In this work, we build a boosted kilonova model with rich physical details, including the description of the evolution and stability of a proto neutron star, and the energy absorption through X-ray photoionization.  We simulate the observation prospects and find the only way to match the absence of detection is to limit the energy injection by the newly born magnetar to only a small fraction of the neutron star rotational energy, thus they should collapse soon after the merger. Our result indicates that most supramassive neutron stars resulting from binary neutron star mergers are short lived and they are likely to be rare in the Universe.
\end{abstract}




\section{Introduction}
Kilonovae (KNe, also called macronovae) are bright optical events that occur after the merger of a binary neutron star (BNS) systems (\citealt{1998ApJ...507L..59L,2010MNRAS.406.2650M}, see \citealt{2015IJMPD..2430012R,2016AdAst2016E...8T,2016ARNPS..66...23F,2019LRR....23....1M} for reviews), serving as the optical counterparts to gravitational wave (GW) sources. They arise from the thermal radiation emitted by the hot matter ejected during the BNS merger. The thermal energy of the ejected material originates from the radioactive decay of heavy elements produced through the r-process nucleosynthesis \citep{1957RvMP...29..547B,1957PASP...69..201C} which happens in a neutron-rich environment. To first order approximation, the evolution of a KN can be treated as an isotropic expanding hot ejecta. The ejecta is initially optically thick due to the bound-bound absorption (i.e., the line forest) by the r-process elements, but gradually gets transparent as it expands, resulting in a peak in the light curve. The spectrum of the emitted radiation, which can be approximated as thermalized emission, typically peaks in the optical or near-infrared wavelengths. Clear KN emission signatures were first observed as an electromagnetic counterpart of the notable event GW170817: the merger of a BNS system detected by the Laser Interferometer Gravitational-Wave Observatory (LIGO) \citep{2017PhRvL.119p1101A, 2017ApJ...848L..12A}. The observations mostly match with the theoretical modeling, and the recognition of Lanthanide elements in the spectrum confirms the r-process heating as the energy source \citep{2017ApJ...848L..17C}. Together with the prompt GRB \citep{2017ApJ...848L..13A}, its afterglow observation, and the host galaxy, GW170817 has been extensively applied in the research of physics and astrophysics, such as the neutron star matter equation of state \citep{2018PhRvL.121p1101A}, GRB afterglow physics \citep[e.g.,][]{GG2018,Lazzati+18,Margutti2018,2019MNRAS.484L..98K,Troja2019,Wu-MacFadyen-19,BGG2020,Nathanail+20,Nakar-Piran-21}, cosmology \citep{2017Natur.551...85A,Hotokezaka+19, 2021ApJ...908..200W} and fundamental physics \citep{2017ApJ...851L..18W}. 

While the brightness of a KN is inherently limited by the radioactive energy of the ejected material (approximately $10^{46}$ erg, e.g. \citealt{2019LRR....23....1M}), there is a possibility of augmenting their luminosity through a hypothesized energy source originating from a central remnant that remains active after the merger event \citep{2013ApJ...776L..40Y,2014MNRAS.439.3916M,2016ApJ...818..104K}. One such example is a millisecond magnetar. If a remnant of the merger persists due to rapid uniform rotation (rigid body rotation), its rotational energy could potentially reach levels up to a few $10^{53}$\,erg \citep{2017ApJ...850L..19M,2018MNRAS.481.3670R} limited by the Keplerian rotation (also known as the mass-shedding limit). At this stage, the neutron star is referred to as a supramassive neutron star (SMNS) since its mass exceeds the maximum allowed mass of a static neutron star, known as the Tolman-Oppenheimer-Volkoff mass ($m_{\rm TOV}$). It is believed that a hypothetical SMNS formed from a BNS merger is likely to also be a millisecond magnetar whose dipole magnetic field ranges from $10^{14}$ to $10^{16}$ G, where the upper limit is bounded by the stability of magnetized NS (e.g., \citealt{2013MNRAS.433.2445A}), and the lower limit is caused by the amplification of magnetic fields during the differential rotation phase of the central remnant following the merger (e.g., \citealt{2006Sci...312..719P}). A millisecond magnetar spins down and losses energy through magnetic dipole radiation. The majority of this released energy is transferred into the surrounding environment by the magnetar wind. If a fraction of this energy can be deposited into the ejecta as thermal energy, it has the potential to significantly enhance the luminosity of a KN - by more than two orders of magnitude \citep{2019LRR....23....1M}, depending on the model. This enhanced luminosity enables detection at distances exceeding that for regular KNe by more than an order of magnitude, corresponding to a detectable volume more than three orders of magnitude greater than that of regular KNe. In this study, we refer to these exceptionally bright, and as of yet hypothetical transients, as magnetar-boosted KNe. Recently, works have argued that their luminosity can be reduced if the ejected material is Poynting flux dominated \citep{2022MNRAS.516.2614A}, or if the ejection is not isotropic \citep{2023arXiv230915141W}. However, such a scenario is not considered in this work since the magnetic fields in the magnetar wind are mostly dissipated in our model (this will be explained in \S \ref{subsec:PWN}). In this paper, since we only care about a magnetar produced after a neutron star merger, we use the terms ``magnetar" and ``SMNS" interchangeably. Readers should not confuse it with the magnetars as remnants of single-star stellar evolution.

Despite the fact that the occurrence rate of  magnetar-boosted KNe may constitute only a small fraction of the overall population of binary neutron star mergers, their detectability can still remain substantial due to the considerably larger detectable volume as compared with regular KNe. Numerous ground-based optical telescopes, such as the Zwicky Transient Facility (ZTF) \citep{2019PASP..131a8002B} and the Panoramic Survey Telescope and Rapid Response System (Pan-STARRS) \citep{2004SPIE.5489...11K}, have been actively surveying the sky for rapidly evolving transients. Additionally, several upcoming optical telescopes, including the Vera C. Rubin Observatory \citep{2019ApJ...873..111I}, are ready to start their operations in the near future. However, over the past several years of sky surveys, no confirmed KNe have been reported \citep{2020ApJ...904..155A, 2021ApJ...918...63A}.

The absence of detection provides a significant constraint on the characteristics and rates of magnetar-boosted KNe, specifically addressing the question of why they are so rare. One potential explanation lies in the formation rate of SMNS. It is possible that the occurrence of long-lasting SMNS is an exceptionally uncommon outcome of BNS mergers. The fate of a BNS merger remnant is determined by factors including the equation of state (EoS) of neutron star matter, the initial rotation speed during uniform rotation (if applicable), and the mass of the remnant. Depending on various conditions, four possible scenarios can arise, ranked here in order of decreasing remnant mass. Firstly, if the remnant is excessively massive, it will promptly collapse into a black hole without undergoing an intermediate stage. Secondly, if the remnant survives sudden collapse, its inner angular momentum will rapidly dissipate and redistribute through differential rotation. At this stage, the remnant is known as a hypermassive neutron star (HMNS). A HMNS may collapse into a black hole if the centrifugal force can't balance the gravity when it slows down. Thirdly, if the remnant remains stable against collapse after it enters uniformly rotating phase, it becomes a temporarily stable SMNS. Lastly, if the remnant's gravitational mass at rest remains below the Tolman-Oppenheimer-Volkoff mass ($m_{\rm TOV}$), it becomes a indefinitely stable neutron star. The boundary between these scenarios relies on the aforementioned conditions, but the EoS and the statistical properties of the remnant's rotation and mass are still not well understood. Considering the lower limit of $m_{\rm TOV}$ constrained by most massive pulsars \citep{2013Sci...340..448A, 2021ApJ...915L..12F}, assuming progenitors follow the mass distribution of Galactic neutron stars, and assuming that SMNS initially rotates at Keplerian speed, the recent work by \cite{2021ApJ...920..109B} suggests that a non-negligible fraction of BNS remnants would result in long-lived SMNS.  Consequently, the absence of detection should place stringent constraints on these assumptions.

Indeed, both observational and theoretical studies have indicated that the long-lived remnants are likely to be very rare. Late time radio observations of sGRBs have so far not shown evidence of a persisting radio source \citep{2014MNRAS.437.1821M}. Recently, \citealt{2021ApJ...920..109B} have found that the long-lived magnetar model are inconsistent with the signatures of X-ray plateaus found in sGRB afterglow, as well as the lack of bright sources in blind radio sources (for the latter point see also earlier predictions by \citealt{2015ApJ...806..224M}). \citealt{2022ApJ...939...51M} performed numerical simulations of neutron star mergers and found that the core of the remnant will collapse into a black hole even if the remnant's total mass and angular momentum allows the formation of a temporarily stable SMNS, since the core slows down much faster than the ``disk". Motivated by these studies, a similar constraint should be made by the aforementioned optical survey, provided that a boosted KNe model is well established.

To accurately predict the signatures of boosted KNe, one needs to carefully study the interaction between the magnetar wind and the ejecta. The energy injection efficiency should be calculated based on the interaction, rather than assuming a free efficiency parameter. A detailed calculation was carried out by \cite{2014MNRAS.439.3916M} (referred to as MP14). They considered the efficiency by incorporating a model involving a pulsar wind nebula (PWN) obstructed by an ejecta wall. In this model, the PWN is inflated by the magnetar wind, while the ejecta wall consists of the r-process elements ejected during the merger. The ejecta is photoionized and heated by the X-rays emitted from PWN.
Within the PWN, ultra-relativistic pairs emit gamma-rays through synchrotron radiation and inverse Compton scattering. The gamma-rays subsequently annihilate with background photons and generate additional ultra-relativistic pairs, initiating the, so called, pair cascade. Due to the small size of the PWN constrained by the ejecta wall, the cascade becomes saturated, resulting in a fraction of $\sim$ 10\% of spindown power turning to the rest mass of pairs in the PWN \citep{1987MNRAS.227..403S}. Consequently, the PWN becomes highly opaque to Thomson scattering, and a significant amount of energy injection eventually turns to the kinetic energy of the ejecta through the pdV work \citep{2014MNRAS.439.3916M}. According to this model, the luminosity enhancement is considerably suppressed as compared with the energy input from the central engine. Nonetheless, they find a magnetar-boosted KN luminosity that is still more than two orders of magnitude brighter than a regular one. Correspondingly, the model predicts a detectable volume that is more than three orders of magnitude larger than that of regular KNe, which is in contrast with the lack of detections. It should be noted that in this model, the assumption is made that the magnetars are indefinitely stable. To further investigate the constraints of the rate of SMNS implied by observations, a more detailed investigation of this model, including the photoionization processes and a limited survival time for SMNS, may be necessary. 

The model can also be improved by considering the Rayleigh-Taylor instability of the PWN-ejecta interacting surface, which arises due to the high acceleration and density difference. If this instability occurs, a significant portion of the matter in the PWN may escape from the ejecta, resulting in the formation of an ultra-relativistic blastwave. This blastwave propagates through the interstellar medium, accelerates electrons, amplifies microscopic magnetic fields, and generates synchrotron radiation, just as in the case of a GRB afterglow. However, unlike a GRB, the blastwave in this scenario is isotropic rather than confined to a narrow jet angle. Considering the substantial energy budget of the SMNS and the isotropic nature of the blastwave, such radiation might also be observed through sky surveys, and would be classified as a, so called, ``orphan afterglow". Such events haven't been robustly identified, further constraining the formation rate of long lived rapidly rotating magnetized NS remnants of BNS mergers.

In this work, we present a refined model of magnetar-boosted KNe building upon the framework established by MP14, incorporating additional physical details. Specifically, we incorporate a limited survival time for the SMNS and a more detailed photoionization calculation of the ejecta. We also explore the potential occurrence of Rayleigh-Taylor instability and its afterglow-like radiation. Using this model, we perform an EoS-independent study to assess the observational potential of such remnants across the parameter space. We also perform a EoS-dependent simulation to study the detection rate, starting from a population of BNS mergers and incorporating the observations made by ground-based optical telescopes. Our results indicate that, in order to be consistent with observations, most SMNS can't be long-lived, suggesting that SMNS, as the merger remnant of BNS, are exceedingly rare in the Universe.

This paper is organized as follows. We describe the details of our model in section \S\ref{sec:Model} and discuss its observational features and stability. In \S \ref{sec:LC} we discuss the observational features and prospects of our model. In \S\ref{sec:observation} we perform the EoS-independent and EoS-dependent study, compare it with current observations, and make constraint on the merger remnants. In \S\ref{sec:implications} we discuss the implications on related topics of our model. Finally, we summarize the points and conclude our study in section \S\ref{sec:summary}. 

\section{Modeling the magnetar-boosted kilonovae}\label{sec:Model}

\begin{figure}
    \centering
    \includegraphics[width=\columnwidth]{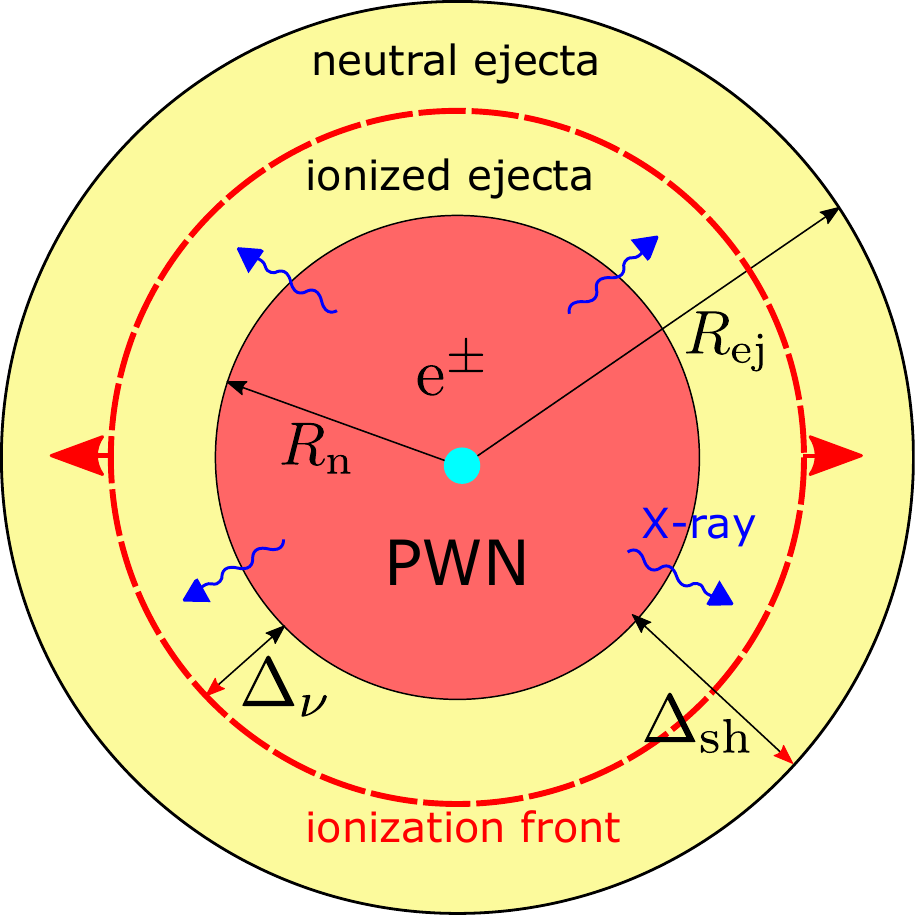}
    \caption{Illustration of the structure of the post-merger system. The system is composed by an inner pulsar wind nebula (PWN) inflated by a magnetar, and an outer ejecta shell. The PWN is composed by X-ray photons and electron-positron pairs. The X-ray radiation ionize and heat the ejecta, leading to a boosted luminosity of KN. The evolution of the X-ray opacity in the ejecta can be characterized by a approximate ionization front (red-dashed line). The X-rays breakout from the ejecta once the ionization front reaches the ejecta surface.}
    \label{fig:structure}
    \label{fig:cartoon}
\end{figure}

In this work, we consider a system consisting of two distinct regions: the inner PWN and the outer ejecta. A schematic representation of the system is illustrated in Figure \ref{fig:cartoon}. The PWN, predominantly composed of electron-positron pairs and X-rays, is inflated by the spindown power of the magnetar. It's surrounded and trapped by the ejecta wall that consists of r-process elements. Initially both PWN and ejecta are optically thick, and most of the internal energy converts to kinetic energy of the ejecta through pdV expansion rather than being radiated away. The x-rays diffuse from the PWN, photoionize and heat the ejecta, and are able to break out once the ejecta is fully ionized. The hot ejecta produces the observed thermal radiation, i.e., the KN. When the magnetar collapses, the PWN loses its energy supply and rapidly disappears due to pair annihilation, leaving an expanding thermal ejecta. We describe the details of the above process in the following parts.

\subsection{The basic assumptions}\label{subsec:assumption}

To simplify the calculation, we build a toy model based on the following assumptions:

(i) The ejecta has a uniform (but time evolving) density $\rho_{\rm ej}$.

(ii) To balance the pressure at the interface between PWN and ejecta, we assume a uniform and radiation dominated pressure throughout the PWN-ejecta system.

(iii) The expansion is homologous. In other words, the velocity $v$ at the radius $r$ follows $v\propto r$.

The uniform pressure and the homologous condition result in following relations
\begin{equation}\label{eq:assumption}
    \left(\frac{E_{\rm n}}{E_{\rm n} + E_{\rm ej}} \right)^{1/3} = \frac{R_{\rm n}}{R_{\rm ej}} = \frac{dR_{\rm n}/dt}{dR_{\rm ej}/dt}
\end{equation}
where $E_{\rm n}$ and $E_{\rm ej}$ are internal energy of nebula and ejecta, respectively. The radii of the PWN $R_{\rm n}$ and the ejecta $R_{\rm ej}$ are measured from the magnetar. Because most of energy is trapped in the PWN due to the high opacity, the above equation implies that $R_{\rm n}\sim R_{\rm ej}$, i.e., the ejecta shell must be thin. In fact, this is a natural result of pressure balance. The shell thickness is $\Delta_{\rm sh}=R_{\rm ej} - R_{\rm n}$.

Assuming a uniform density $\rho_{\rm ej}$ in the ejecta, the kinetic energy of the ejecta is
\begin{align}\label{eq:kinetic}
    E_{\rm k} &= \int_{R_{\rm n}}^{R_{\rm ej}} \frac{1}{2}4\pi r^2\rho_{\rm ej}v^2dr \nonumber \\
        &= \frac{3}{10}\left(\frac{dR_{\rm ej}}{dt}\right)^2\frac{1-(R_{\rm n}/R_{\rm ej})^5}{1-(R_{\rm n}/R_{\rm ej})^3}m_{\rm ej},
\end{align}
where $m_{\rm ej}$ is the mass of the ejecta. The kinetic energy of the PWN is not considered because it's much lighter than ejecta. We also don't consider relativistic effects for the bulk motions in our model. In an extreme case, if the magnetar has a rotational energy $E_{\rm ini}=10^{53}$ erg and the ejecta has a mass of $m_{\rm ej}=0.01M_{\odot}$, the ejecta is accelerated to mildly relativistic speed. However, we don't expect a significant modification to our results because: (i) In the case a SMNS forms, most of the ejecta mass comes from disk wind ejecta with a mass of approximately $0.1 M_{\odot}$ (e.g., \cite{2019ApJ...880L..15M}), which cannot be easily accelerated to relativistic speeds. (ii) The rotational energy is generally expected to be $E_{\rm k} \lesssim 10^{53}$ erg constrained by the Keplerian rotation.

For the readers' convenience, we list the symbols of geometric and thermodynamic variables in Table \ref{tab:geo_and_therm_symbols}.

\begin{table}
	\centering
	\caption{Thermaldynamic variables.}
	\label{tab:geo_and_therm_symbols}
	\begin{tabular}{cc} 
		\hline
		Symbol & Definition\\
		\hline
		  $E_{\rm n}$ & Internal energy of the PWN\\
		$E_{\rm ej}$ & Internal energy of the ejecta\\
            $E_{\rm k}$ & Kinetic energy of the ejecta \\
            $R_{\rm n}$ & Radius of the PWN (measured from the magnetar) \\
            $R_{\rm ej}$ & Radius of the ejecta (measured from the magnetar) \\
            $V_{\rm tot}$ & Volume of the PWN + ejecta \\
            $V_{\rm n}$ & Volume of the PWN \\
            $\Delta_{\rm sh}$ & Thickness of the ejecta shell \\
            $\rho_{\rm ej}$ & Density of the ejecta \\
		\hline
	\end{tabular}
\end{table}

\subsection{Magnetar}
Assuming magnetic dipole radiation, the spindown power of the magnetar is
\begin{equation}\label{eq:spindown}
    L_{\rm sd}(t)=L_{\rm sd,0}(1+t/t_{\rm sd})^{-2}.
\end{equation}
Note the spindown power index may not be precisely -2 when considering the variation  of the moment of inertia during the early evolution of a fast spinning SMNS. However, this approximation provides a reasonable approximation of the spindown process.
The initial spindown power is assumed to follow the magnetic dipole power
\begin{equation}
    L_{\rm sd,0} = \frac{B^2R_{\rm NS}^6\Omega_0^4}{2c^3},
\end{equation}
which depends on the magnetar dipolar magnetic field $B$ at the pole, the initial angular velocity $\Omega_0$ and the neutron star radius $R_{\rm NS}$. In this work we parameterize the power by the initial rotational energy $E_{\rm ini}$ instead of the angular velocity, because this determines the energy budget available to the KN system. $E_{\rm ini}$ is given by $E_{\rm ini}=\frac{1}{2}I\Omega_0^2$ with the moment of inertia estimated by $I=\frac{2}{5}M_{\rm NS}R_{\rm NS}^2$. Assuming $M_{\rm NS}\approx 3M_{\odot}$ and $R_{\rm NS}\approx 10$ km, the initial spindown power becomes
\begin{equation}\label{eq:initial_spindown}
    L_{\rm sd,0}\approx10^{50}\left(\frac{E_{\rm ini}}{10^{53}\mathrm{erg}}\right)^2\left(\frac{B}{10^{15}\mathrm{G}}\right)^2{\rm erg/s}.
\end{equation}

Provided the initial rotational energy $E_{\rm ini}$, the spin-down time scale $t_{\rm sd}$ can be calculated by $t_{\rm sd}=E_{\rm ini}/L_{\rm sd,0}$. Inserting the above equations, we have
\begin{equation}\label{eq:spindown_time}
    t_{\rm sd} = 0.01\left(\frac{E_{\rm ini}}{10^{53}\mathrm{erg}}\right)^{-1}\left(\frac{B}{10^{15}\mathrm{G}}\right)^{-2}\ {\rm d}.
\end{equation}

If the magnetar is indefinitely stable, or if the spin-down timescale is shorter than the KN peak time, its entire rotational energy is extracted and becomes available for enhancing the KN. However, in cases where the initial rotation speed at the onset of the uniform rotation stage is significantly smaller than the Keplerian speed, or if the magnetar is so massive that it collapses on a finite timescale $t_{\rm c}$, only a fraction of the energy will be accessible for the KN. We define this fraction as the energy extraction ratio $\eta$. Its value is calculated as follows: 
\begin{equation}
    \eta=\frac{\int_0^{t_{\rm c}}{L_{\rm sd}dt}}{E_{\rm ini}}=\frac{t_{\rm c}}{t_{\rm c}+t_{\rm sd}}.
\end{equation}

The above discussion leads to a parametrization of spindown luminosity by the magnetic field $B$, initial rotational energy $E_{\rm ini}$, and the energy extraction ratio $\eta$. We list these symbols related with the magnetar in Table \ref{tab:magnetar_symbols}.

\begin{table}
	\centering
	\caption{Magnetar variables.}
	\label{tab:magnetar_symbols}
	\begin{tabular}{cc} 
		\hline
		Symbol & Definition\\
		\hline
		  $L_{\rm sd}$ & Spindown power\\
		$L_{\rm sd,0}$ & Initial spindown power\\
            $t_{\rm sd}$ & Spindown timescale \\
            $E_{\rm ini}$ & Initial rotational energy of the magnetar \\
            $B$ & Magnetic field at the magnetar poles \\
            $t_{\rm c}$ & Magnetar survival timescale \\
            $\eta$ & Energy extraction ratio of the magnetar \\
		\hline
	\end{tabular}
\end{table}

\subsection{PWN}\label{subsec:PWN}

The magnetar continuously injects ultra-relativistic pairs ($\gamma \gg 10^4$) into the system, leading to the formation of a PWN. The pairs quickly cool down through synchrotron radiation and inverse Compton scattering, producing high energy gamma-ray photons. The gamma-ray photons are able to annihilate with background photons which further produces ultra-relativistic pairs, resulting in a pair cascade. The extent of this cascade is characterised by the compactness parameter $\ell \equiv \sigma_{\rm T} L_{\rm sd}/(R_{\rm n} m_{\rm e} c^3)$. When $\ell \gg 1$, the pair cascade is saturated. As a result, around $Y\approx 0.1$ of spindown power turns to rest mass energy of the pairs. Consequently, the PWN is predominantly composed of low-energy electrons and non-thermal photons. The spectrum of the photons extends from the background photon energy to the pair annihilation threshold $\sim$ 1 MeV with a power-law index of -1 (\citealt{1987MNRAS.227..403S, 2013LNP...873.....G}; see \citealt{2021ApJ...917...77V} for a more detailed calculation in the case of superluminous supernova). Since the PWN is trapped behind the ejecta wall, in this work, we approximate the background photon energy by the typical thermal photons energy of the ejecta, i.e., $3k_{B}T_{\rm ej}$, where $k_{B}$ is the Boltzmann constant and $T_{\rm ej}$ is the temperature of the ejecta. Note that because the energy density of the system is uniform, the PWN and the ejecta should share a common temperature
\begin{equation}
    T_{\rm ej} = \left(\frac{E_{\rm n} + E_{\rm ej}}{aV_{\rm tot}}\right)^{1/4},
\end{equation}
where $a$ is the Stefan–Boltzmann constant.

The pair density $n_{\pm}$ (counting both electrons and positrons) is estimated by balancing the pair production and pair annihilation rate. The pair production rate can be estimated by saturated pair cascade $\dot{n}_+=YL_{\rm sd}/(m_{\rm e}c^2V_{\rm n})$. Balancing with the pair annihilation rate $\dot{n}_-=3\sigma_{\rm T}cn_{\pm}^2/16$, the pair density is calculated by 
\begin{equation}\label{eq:pair density}
    n_{\pm}=\sqrt{\frac{16YL_{\rm sd}}{3\sigma_{\rm T}m_{\rm e}c^3V_{\rm n}}}.
\end{equation}
The timescale to reach this equilibrium is $t_{eq}\simeq 16R_{\rm n} /3c\tau_{\rm n}$ \citep{2014MNRAS.439.3916M}, which is typically shorter than the evolution timescale. We have tried to incorporate a dynamical pair density considering $dn_{\pm}/dt=\dot{n}_+ - \dot{n}_-$, but we find no practical difference as compared with the balanced value.

The radiation of the PWN can be estimated using the photon diffusion timescale $t_{\rm n}^{\rm d} = R_{\rm n}(1+\tau_{\rm n})/c$, where the optical depth $\tau_{\rm n}$ caused by Thomson scattering can be estimated by
\begin{equation}\label{eq:pair optical depth}
    \tau_{\rm n} = n_{\pm}\sigma_{\rm T}R_{\rm n}=\sqrt{\frac{4Y\sigma_{\rm T}L_{\rm sd}}{\pi R_{\rm n}m_{\rm e}c^3}}.
\end{equation}
The diffusion timescale, $t_{\rm n}^{\rm d}$, is a smooth interpolation between the optically thin and thick cases. Now the luminosity of the PWN can be estimated by
\begin{equation}
    L_{\rm n}\sim\frac{E_{\rm n}}{t^d_n}=\frac{cE_{\rm n}}{R_{\rm n}(1+\tau_{\rm n})}.
\end{equation}

Considering the shape of the spectrum, we can estimate the frequency dependent luminosity
\begin{equation}
    L_{\rm n,\nu}=\frac{L_{\rm n}}{h\nu\int_{3k_{\rm B}T_{\rm ej}}^{1MeV}\epsilon^{-1}d\epsilon}.
\end{equation}

After the survival timescale $t_{\rm c}$, the magnetar collapses into a black hole. The sudden termination of the ultra-relativistic pair supply ceases the pair cascade. However, it may take some time before this information propagates to the nebula surface, whose speed can be approximated by the sound speed. 
Since the PWN material is a relativistic fluid, the sound speed is $c_s=c/\sqrt{3}$, which can only be well-modeled by considering the relativistic effects. This is out of the scope of our work. To simplify the scenario, we simply assume that this information is instantaneously transmitted across the PWN, so the pair density directly starts dropping following the annihilation rate. At the same time, we also turn the spindown power $L_{\rm sd}$ to 0. In this approximation the PWN quickly disappears after the collapse because it quickly becomes transparent. Although our treatment exaggerates the effects of the collapse, it is unlikely to strongly affect the peak luminosity, because $t_{\rm c}$ is generally much earlier than the time at which the ejecta becomes transparent, at which stage the physical evolution of the PWN is hidden by the surrounding ejecta wall. From the observer's perspective, the central engine is active for a very short timescale, so the ejecta appears as if it undergoes an instantaneous energy injection, where the specific details of the injection process are no longer important.

The symbols related with the radiation of PWN (as well as ejecta to be discussed next) are listed in Table \ref{tab:rad_symbols}.

\begin{table}
	\centering
	\caption{Radiation process variables.}
	\label{tab:rad_symbols}
	\begin{tabular}{cc} 
 
		\hline
		Symbol & Definition\\
		\hline
		$n_{\pm}$ & Total number density of electrons and positrons\\
            $n^{\rm i}$ & number density of ions at $i^{\rm th}$ ionization state \\
            $n_{\rm e}$ & number density of free electrons in the ejecta \\
            $\tau_{\rm n}$ & Optical depth of the PWN \\
            $\tau_{\rm ej}$ & Optical depth of the ejecta (optical and near infrared) \\
            $\kappa_{\rm abs,\nu}$ & Absorption optical depth of the ejecta (X-ray) \\
            $\kappa_{\rm es,\nu}$ & scattering optical depth of the ejecta \\
            $\Delta_{\nu}$ & X-ray penetration depth of the ejecta \\
            $A^{\rm ref}_{\nu}$ & Reflection rate of the X-rays \\
            $A^{\rm abs}_{\nu}$ & Absorption rate of the X-rays \\
            $A^{\rm tra}_{\nu}$ & Transmission rate of the X-rays \\
            $T_{\rm ej}$ & Temperature of ejecta \\
            $T_{\rm eff}$ & Effective temperature of ejecta \\
            $L_{\rm n}$ & Luminosity of PWN \\
            $L_{\rm n,\nu}$ & Frequency-dependent luminosity of PWN \\
            $L_{th}$ & Thermal radiation luminosity of the ejecta \\
            $F_{\nu}$ & Flux of the thermal radiation\\
		\hline
	\end{tabular}
\end{table}

\subsection{Ejecta}\label{subsec:ejecta}
The composition of the ejecta is rather complicated and is subject to numerical study. The main challenge here is the modelling of photoionization of the ejecta, which requires the knowledge of the bound-free cross sections of r-process elements. However, due to the relatively short half-decay timescales of these elements and their isotopes, it is difficult to measure these values in ground-based laboratories. Currently, the available atomic data for the heaviest elements is iron-56. Moreover, in the situation where a long-lived magnetar is present, it will strongly irradiate the disk outflows with neutrinos, which tends to increase the electron fraction of the material to values $\gtrsim$ 0.3 \citep{2017MNRAS.472..904L}. As a result, the ejecta will be mostly composed by light r-process elements whose electron structure is similar to iron-56. Therefore, in this work, we follow MP14 and assume the ejecta is iron like. While this is a crude estimation, we will demonstrate in the following section that the process of thermalization does not play a dominant role in determining the luminosity of magnetar-boosted KNe.

Similar to PWN, the ejecta is initially optically thick. The radiation in the X-ray band suffers bound-free absorption, and the optical rays suffer bound-bound absorption. The heating efficiency of the ejecta and its resulting luminosity sensitively depend on the photoionization process, which we will discuss in detail below. Similar to previous sections, we summarize the frequently used symbols in Table \ref{tab:rad_symbols}.

\subsubsection{Ionization}\label{subsubsec:ionization}
The X-rays radiated into the ejecta lead to the photoionization of the elements. As mentioned above, we approximate the ejecta by matter composed of iron-56 which is initially neutral. The ionization is assumed to be balanced throughout the whole time and evolves as a quasi-static process. The ionization balance in a photon bath is 
\begin{equation}\label{eq:ionization-ori}
    n^i\int \frac{4\pi J_{\nu}}{h\nu}\sigma^i_{\nu}d\nu = \alpha_{\rm rec}^i n^{i+1}n_{\rm e},
\end{equation}
where $n^i$ is the number density of the $i^{th}$ excited state of iron, $J_{\nu}$ is the radiative intensity of the X-rays, $\sigma_{\nu}^i$ is the photoionization cross section, $\alpha^i_{\rm rec}$ is the recombination rate \citep{1981ApJ...249..399W} which depends on ejecta temperature $T_{\rm ej}$, and $n_{\rm e}$ is the number density of free electrons. In our definition, $i$ starts from 1 to 27 for iron, where 1 correspond to neutral iron and 27 corresponds to fully ionized iron. The approximations of cross section for different ions are picked from \cite{1995A&AS..109..125V} and \cite{1996ApJ...465..487V}, which are analytical interpolations of atom data. In principle, The above approximations are effective from the ionization threshold energy and up to 0.5MeV, above which relativistic corrections need to be considered. Here we simply extrapolate them to 1MeV, assuming the relativistic effects don't strongly impact our overall results. The relation between cross section and photon energy plays a crucial role in calculating energy absorption and X-ray radiation, as it directly determines the optical depth of X-rays. Roughly speaking, the cross section peaks at the threshold energy and decreases approximately following a power-law relation. The threshold energy tends to be higher for high excited states compared to low excited states, while the cross section at the threshold energy tends to be lower at high excited states. This is due to the difficulty of ionizing inner shell electrons that have greater binding energies than outer shell electrons. 

Because the ejecta is initially neutral, all free electrons come from ionized atoms, so we have
\begin{equation}
    n_{\rm e}=\sum_{i=1}^{i=27}(i-1)n^i.
\end{equation}
The X-ray intensity can be estimated by the luminosity of the PWN at the interface
\begin{equation}
    4\pi J_{\nu}=\frac{L_{\rm n,\nu}}{4\pi R_{\rm n}^2}.
\end{equation}

We normalize the ion number density to $f^i=n^i/n_{\rm Fe}$ where $n_{\rm Fe}=\rho_{\rm ej}/(56m_{\rm p})$ is the iron atom number density. The degree of ionization can be expressed by the fraction of free electrons to total electrons, i.e., $f_{\rm e}=n_{\rm e}/(26n_{\rm Fe})$, or $1-f_{\rm e}$ which is a measure of the optical depth of the ejecta. Now $f^i$ can be solved by the following equations
\begin{align}
    & f^i\int \frac{4\pi J_{\nu}}{h\nu}\sigma^i_{\nu}d\nu = n_{\rm Fe}\alpha_{\rm rec}^i f^{i+1}\sum_{i=1}^{i=27}(i-1)f^i, \\
    & \sum_{i=1}^{i=27}f^i=1.
\end{align}
Given the solutions, we can calculate the bound-free opacity
\begin{equation}
    \kappa_{\rm bf,\nu} = \frac{1}{56m_{\rm p}}\sum_{i=1}^{26}f^i\sigma_{\nu}^i
\end{equation}

In addition to bound-free absorption, the hard X-rays can also be absorbed by down scattering, the corresponding effective opacity can be estimated by
\begin{equation}
    \kappa_{\rm es,\nu}=\kappa_{\rm es}\frac{h\nu}{m_{\rm e}c^2},
\end{equation}
where the scattering opacity is $\kappa_{\rm es}=26\sigma_{\rm T}/(56m_{\rm p})\approx 0.2\mathrm{cm^2g^{-1}}$. Note that X-ray photons can also be scattered by bound electrons. 

The total X-ray opacity of the ejecta is then
\begin{equation}
    \kappa_{\rm abs,\nu}=\kappa_{\rm bf,\nu} + \kappa_{\rm es,\nu}.
\end{equation}

In MP14, the propagation of photoionization is approximated by an ionization front at which optical depth is equal to 1. X-rays are not allowed to escape from the ejecta until the ionization front reaches the ejecta surface. The ionization fronts are associated with the ion species which dominate the photoionization. In contrast to this approach, we consider a different way of modeling the ionization. Using the opacity estimated above, we are able to track the optical depth at each frequency.
As an X-ray photon traverses through the ejecta, it can either pass through, be absorbed through ionization, or reflected back due to scattering. The fate of this X-ray photon depends on the optical depth of absorption and scattering at its frequency. Thus, we do not employ an assumption of ionization front, but rather a frequency dependent transmission rate. Our approach does not deviate significantly from the "front" approach, but it results in a smoother X-ray light curve during the breakout time. In addition, since the transmission rate is frequency-dependent rather than associated with specific ion species, we are able to model the evolution of the X-ray spectrum. The details of this calculation will be shown in \S\ref{sec:scattering}.

Although our model does not require an ionization front, we can still define an effective penetration depth using a similar approach to MP14. This depth can be considered as an indicator of the transparency of X-ray at a specific frequency. When this depth reaches the surface it provides an estimate for the time of X-ray breakout. 

Following MP14, we approximate the depth by assuming the effective optical depth is equal to 1. The effective optical depth $\tau_{\rm eff,\nu}$ is the absorption optical depth corrected by the path length factor due to scattering
\begin{equation}\label{eq:effective_tau}
    \tau_{\rm eff,\nu} = \tau_{\rm abs,\nu} (1 + \tau_{\rm es,\nu}).
\end{equation}
The absorption and scattering optical depth can be calculated by
\begin{align}
    &\tau_{\rm abs,\nu}=\rho_{\rm ej}\kappa_{\rm abs,\nu}\Delta_{\nu}\\
    &\tau_{\rm es,\nu}=\rho_{\rm ej}\kappa_{\rm es}\Delta_{\nu},
\end{align}
where $\Delta_{\nu}$ is the frequency dependent penetration depth. Equating eq. \ref{eq:effective_tau} to unity, this value can be analytically solved
\begin{equation}
    \Delta_{\nu}=\frac{-1+\sqrt{1+4\kappa_{\rm es}/\kappa_{\rm abs,\nu}}}{2\rho_{\rm ej}\kappa_{\rm es}}.
\end{equation}

It's maximum value is limited to the ejecta thickness $\Delta_{\rm sh}$. 

\subsubsection{Scattering}\label{sec:scattering}

Because of the scattering effect, X-ray photons not only have the possibility of being absorbed or passing through the ejecta but also a chance of being reflected back to the PWN. As mentioned above, the overall effect can be described by frequency-dependent rates of reflection ($A^{\rm ref}_{\nu}$), absorption ($A^{\rm abs}_{\nu}$), and transmission ($A^{\rm tra}_{\nu}$), which satisfy the normalization $A^{\rm ref}_{\nu} + A^{\rm abs}_{\nu} + A^{\rm tra}_{\nu} = 1$. The values of these rates directly depends on the frequency dependent absorption optical depth ($\tau_{\rm abs}$) and scattering optical depth ($\tau_{\rm sca}$). However, their relation is very complicated, as the joint process of absorption and scattering is highly non-linear. The only way to determine the relation is through a Monte Carlo simulation. Unlike the approach taken by MP14, who only simulated cases where the ejecta is optically thick due to both scattering and absorption (resulting in a dependence solely on $\tau_{\rm abs}/\tau_{\rm sca}$), we aim to cover all possible combinations of $\tau_{\rm abs}$ and $\tau_{\rm sca}$. The simulation is described as follows.

\begin{figure}
    \centering
    \includegraphics[width=\columnwidth]{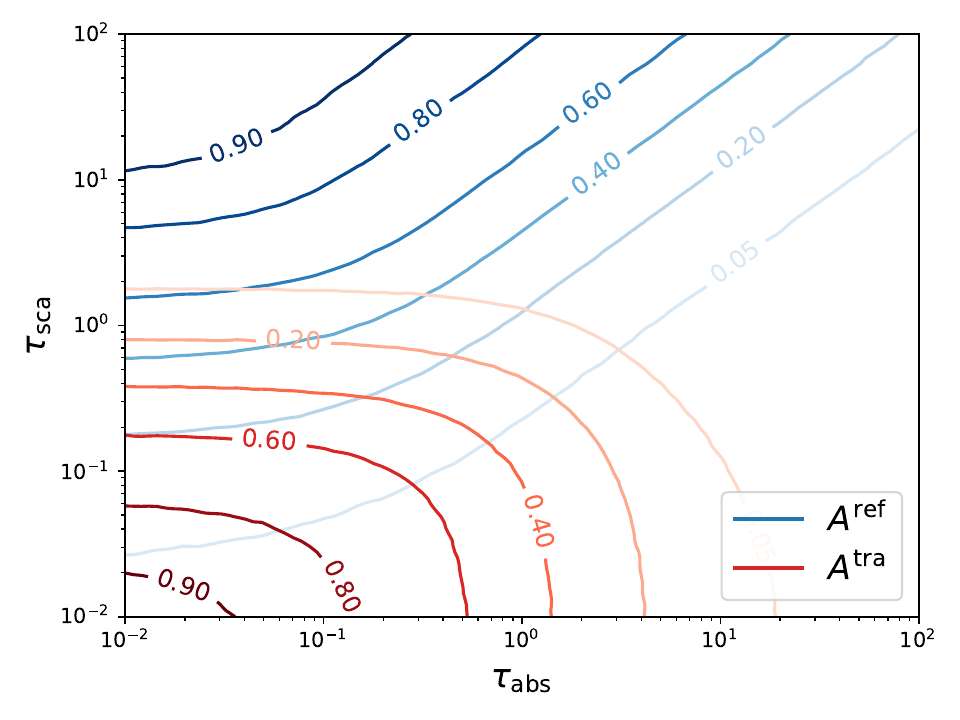}
    \caption{The Monte Carlo simulation result of absorption, reflection, and transmission rate of a slab caused by both absorption and scattering processes. The red lines show the contour plot of $A^{\rm tra}$ as a function of optical depth of two processes, and the blue lines are the $A^{\rm ref}$. $A^{\rm abs}$ can be calculated by $1-A^{\rm tra}-A^{\rm ref}$.}
    \label{fig:albedo}
\end{figure}

We consider a slab with a width of unity and an infinite area. The normal direction of the slab is represented by the z-axis. The slab has an optical depth of $\tau_{\rm abs}$ due to absorption and $\tau_{\rm sca}$ due to scattering. Before injecting a photon, we generate a random variable, $l_{abs}$, which represents the maximum path length the photon can travel before being absorbed. The probability of a photon being absorbed after traveling an accumulated path length $l$ follows a distribution depending on the mean free path $\Bar{l}$: $p(l)=\exp(-l/\Bar{l})/\Bar{l}$. In our setup, the mean free path is $1/\tau_{\rm abs}$. So $l_{abs}$ follows the probability distribution $p(l)=\tau_{\rm abs}\exp(-l\tau_{\rm abs})$. After generating this variable, we inject photons from one side with random directions, but with a positive z-component of velocity. These photons then start a 3D random walk within the slab due to scattering. Since scattering can be regarded as an absorption and emission process, the length of each step follows the same probability distribution with mean free path $1/\tau_{\rm sca}$: $p(l)=\tau_{\rm sca}\exp(-l\tau_{\rm sca})$. We terminate the photon's walk if (i) the photon is absorbed, i.e., the cumulative path length exceeds $l_{abs}$, (ii) the photon is reflected, i.e., $z<0$, or (iii) the photon passes through the slab, i.e., $z>1$. For every pair of $(\tau_{\rm abs}, \tau_{\rm sca})$, we inject one million photons and calculate the reflection, absorption, and transmission rates. These rates, as functions of $\tau_{\rm abs}$ and $\tau_{\rm sca}$, are presented in a contour plot shown in Fig. \ref{fig:albedo}. The results are then interpolated to obtain a smooth function.

\subsubsection{Thermal Radiation}\label{subsubsec:thermal}
Similar to \S\ref{subsec:PWN}, the thermal luminosity of ejecta can be estimated by
\begin{equation}
    L_{\rm th}=\frac{cE_{\rm ej}}{R_{\rm ej}(1+\tau_{\rm ej})},
\end{equation}
where $\tau_{\rm ej}=\rho_{\rm ej}\kappa_{\rm ej}\Delta_{\rm sh}$ is the bound-bound optical depth. The opacity $\kappa_{\rm ej}$ is hard to estimate due to the poor knowledge of the line forest \citep{2019LRR....23....1M}. As previously mentioned, the ejecta is likely composed of iron-like elements with an electron fraction $Y_{\rm e}\gtrsim 0.3$ \citep{2017MNRAS.472..904L}. Such a composition likely results in a relatively low opacity \citep{2020MNRAS.496.1369T}. Here we follow the common simplification in literature and set it to a constant value $\kappa_{\rm ej}=1\mathrm{cm^{-2}g^{-1}}$ (though we consider a broader parameter space in \S \ref{subsec:MC}).

We assume that all photons absorbed are turned to thermal energy, and assume that the ejecta is in thermal equilibrium. The radiation then follows the black body formula. The black body temperature of the ejecta is the effective temperature at the surface
\begin{equation}
    T_{\rm eff}=\left(\frac{L_{\rm th}}{4\pi\sigma R_{\rm ej}^2}\right)^{1/4}.
\end{equation}

The flux of the KN is
\begin{equation}\label{eq:flux}
    F_{\nu}=(1+z)\pi B_{(1+z)\nu}(R_{\rm ej}/D_{L})^2,
\end{equation}
where $B_{\nu}$ is Planck's formula
\begin{equation}
    B_{\nu} = \frac{2h\nu^3}{c^2}\frac{1}{e^{h\nu/k_{\rm B} T_{\rm eff}}-1}
\end{equation}
and $D_{\rm L}$ is the luminosity distance. We have considered cosmological effects here, since later we will show that the most optimal magnetar-boosted KNe can be observed up to a few Gpc. In this study we assume the following cosmology parameters: $H_0=68$ km s$^{-1}$Mpc$^{-1}$, $\Omega_m=0.286$ and $\Omega_{\Lambda}=0.714$.

\subsection{Evolution Equations}\label{subsec:equtions}
Now that we have all the necessary ingredients, we are ready to derive the evolution equations. The system losses its internal energy due to pdV work and radiation. Applying Eq. \ref{eq:assumption} and Eq. \ref{eq:kinetic}, the evolution equations can be summarized as follows
\begin{align}
    \frac{dE_{\rm n}}{dt} & = -\frac{E_{\rm n}}{R_{\rm ej}}\frac{dR_{\rm ej}}{dt} - \int(1-A^{\rm ref}_{\nu})L_{\rm n,\nu}d\nu + L_{\rm sd},\\
    \frac{dE_{\rm ej}}{dt} & = -\frac{E_{\rm ej}}{R_{\rm ej}}\frac{dR_{\rm ej}}{dt}+\int A^{\rm abs}_{\nu}L_{\rm n,\nu}d\nu - L_{th},\\
    \frac{dE_{\rm k}}{dt} &= \frac{E_{\rm n}+E_{\rm ej}}{R_{\rm ej}}\frac{dR_{\rm ej}}{dt},\\
    \frac{dR_{\rm ej}}{dt} & = \left(\frac{10E_{\rm k}}{3m_{\rm ej}}\right)^{1/2}\left[\frac{1-(R_{\rm n}/R_{\rm ej})^3}{1-(R_{\rm n}/R_{\rm ej})^5}\right]^{1/2}
\end{align}

The system can be solved given the parameter set 
$B$, $E_{\rm ini}$, $\eta$, $m_{\rm ej}$. In our calculations, the ejecta is initially taken to be a sufficiently small spheroid with an initial velocity of 0.1c. The results are independent of the initial ejecta radius.

In this study we omit the r-process heating of the ejecta. This is because the radioactive power (e.g., \citealt{2012MNRAS.426.1940K}) is many orders of magnitudes smaller than the dipole power of the magnetar, and has no practical effect on our results. 

\subsection{Rayleigh-Taylor Instability}
While solving the evolution equations, we also test the Rayleigh-Taylor (RT) instability of the system. This is very likely to occur in this scenario, because the heavy ejecta is accelerated by the light PWN matter at early times. If the ejecta breaks apart before the light curve peaks, the PWN matter will leak away forming an ultra-relativistic blastwave, and the rest of the energy will be insufficient to boost the KN. Fully capturing the dynamics of the RT instability would require hydrodynamical simulations, which is beyond the scope of our work. Here we simply consider the linear growth rates and provide a rough test.

The growth timescale of RT instability is roughly estimated by
\begin{equation}
    t_{\rm RT} = 1/\sqrt{Ag\alpha},
\end{equation}
where A is the Atwood number
\begin{equation}
    A=\frac{\rho_{\rm ej} c^2-E_{\rm n}/V_{\rm n}}{\rho_{\rm ej} c^2+E_{\rm n}/V_{\rm n}},
\end{equation}
where $g$ is the acceleration of the ejecta
\begin{equation}
    g \approx \frac{dE_{\rm k}/dt}{m_{\rm ej}dR_{\rm ej}/dt} = \frac{E_{\rm n} + E_{\rm ej}}{m_{\rm ej}R_{\rm ej}}
\end{equation}
and $\alpha$ is the wave number of the instability. Here we approximate it by $\alpha\approx 2\pi/\Delta_{\rm sh}$. 

We calculate this value throughout the evolution of the system. The system is considered unstable if the growth timescale is shorter than the dynamic timescale, i.e., $t_{\rm RT} < t_{\rm dyn}$. The dynamic timescale is defines to be $t_{\rm dyn} = R_{\rm ej}/\beta_{\rm ej}c$, where $\beta_{\rm ej}c = dR_{\rm ej}/dt$ is the velocity of the ejecta. The evolution of the factor $t_{\rm RT}/t_{\rm dyn}$ will be shown together with other variables in \S \ref{sec:LC}, where we will see that the system is generally {\it unstable} to RT instability.

\section{Light Curve Signatures}\label{sec:LC}
In this section, we investigate the behavior of the PWN-ejecta system over the parameter space and the predicted optical and X-ray light curves. 

\begin{figure}
    \centering
    \includegraphics[width=\columnwidth]{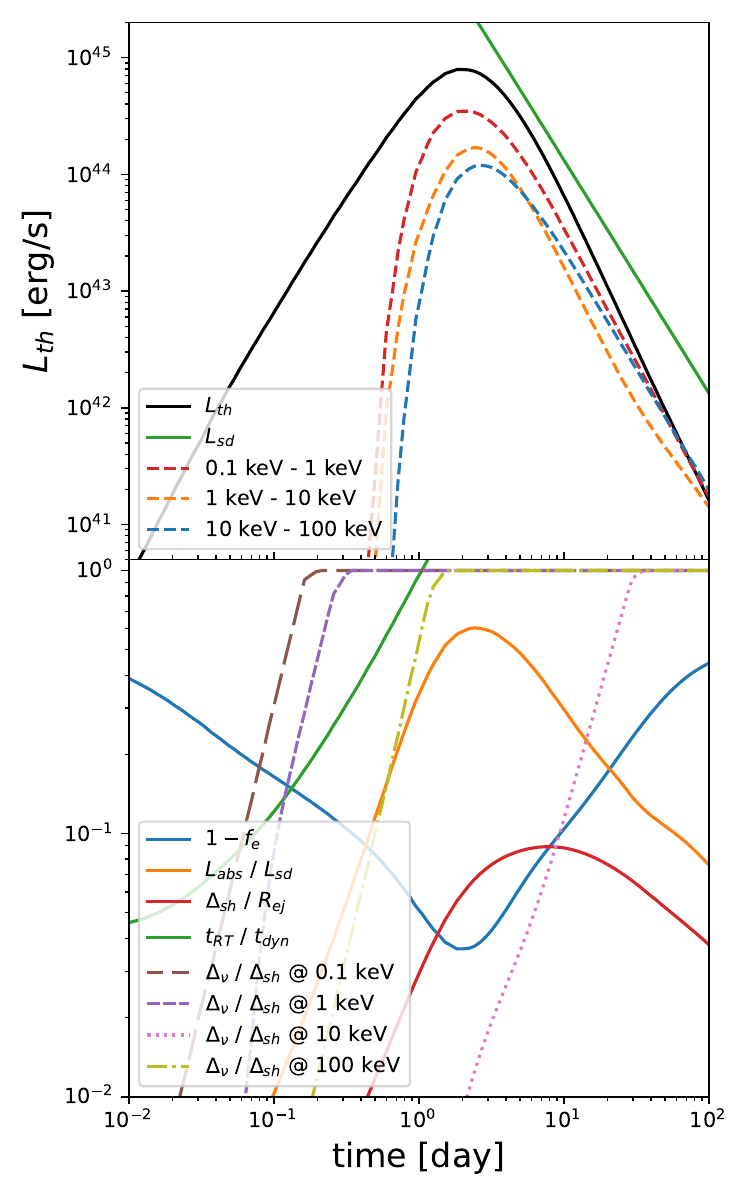}
    \caption{The solution of a typical parameter set: $B=10^{15}$ G, $E_{\rm ini}=3\times10^{52}$ erg, $\eta$ = 1, $m_{\rm ej} = 0.1$ $M_{\odot}$. In the upper panel we show the bolometric luminosity of the optical radiation by the black solid line, and the X-ray luminosity in given bands by colored dashed lines. Together with the luminosity, we also show the spin down power in a green solid line for reference. In the lower panel, the solid colored lines show the evolution of some typical parameters, including the reverse of ionization degree $1 - f_{\rm e}$ (an indicator of X-ray opacity), heating efficiency $L_{\rm abs}/L_{\rm sd}$, the growth rate of RT instability $t_{\rm RT}/t_{\rm dyn}$, and the ejecta shell thickness $\Delta_{sd}/R_{\rm ej}$ as indicated in the figure. The dashed and dotted colored lines show the evolution of the approximate ionization front of X-rays at typical frequencies.}
    \label{fig:default}
\end{figure}

\subsection{Typical temporal behavior of a indefinitely stable kilonova}

First, in Figure \ref{fig:default}, we show the temporal evolution of some critical parameters in an indefinitely stable magnetar boosted KN. The parameters are described in the figure caption. The upper panel shows the luminosity of thermal (optical) and X-ray radiation. The KN (optical) peaks after a few days with a luminosity $\sim 10^{45}$ erg/s. The X-rays show a (frequency-dependent) sharp breakout as expected. The peak luminosity of the X-rays is rather complicated, since it sensitively depends on the photoionization process. Our results are in general agreement with MP14. In MP14, the X-ray trend after the peak follows the same power index as the magnetic dipole formula. This is because, at that stage, the ejecta is fully ionized, allowing X-rays from the PWN to freely escape. However, in our model, with the exception of the relatively low mass of the ejecta, we find that the X-ray power index is initially slightly steeper than the dipole radiation. This is caused by the recombination of ions that increases the optical depth of ejecta, as demonstrated in the lower panel of the figure. However, at an even later stage, when $1-f_{\rm e}$ becomes constant, we still anticipate the same asymptotic trend following the magnetar spindown power.

In the lower panel of the figure, we present the evolution of several key variables of the system. The $1-f_{\rm e}$ term represents the degree of ionization, which serves as an indicator of the X-ray opacity. It can be seen that the ejecta is highly ionized around the peak of the light curve, indicating that the radiation of the PWN is sufficient to fully ionize the ejecta, allowing X-rays to pass through. This result is compatible with previous studies using photoionization code \texttt{CLOUDY} (e.g., \citealt{2018MNRAS.481.2407M}). As the X-ray intensity decreases after the peak, the ionization degree decreases and the opacity increases due to recombination. The heating efficiency, characterized by $L_{\rm abs}/L_{\rm sd}$, varies significantly throughout the evolution, ranging from 0.01 to 0.5. Unlike some studies that assume a constant value, we find that the heating efficiency is dynamic in our model. The evolution of the variable $\Delta_{\rm sh}/R_{\rm ej}$ confirms that the ejecta is compressed into a thin shell due to the high pressure of the PWN. The test of the Rayleigh-Taylor (RT) instability is shown by the ratio $t_{\rm RT} / t_{\rm dyn}$. This ratio is generally less than 1 in the early stages, indicating that the system is prone to RT instability. It is important to note that our modeling only accounts for the linear stages of the RT instability, and may not fully capture the entire process. Numerical hydrodynamics is needed for a more accurate study, but is out of the scope of our work. Furthermore, even if the KN is disrupted by the instability, the existence of a SMNS can be revealed by the non-thermal signatures of the resulting blastwave. We will explain this case in \S \ref{sec:blast-wave}. 

\subsection{Typical features of optical radiation}\label{subsec:lc features}

The most important features of an optical light curve are its peak luminosity (or flux) and peak time. The peak luminosity is difficult to analytically estimate in our model because it depends on the details of ionization, but the peak time is relatively easier. Before showing our numerical result, we first provide an analytical estimation of peak time $t_{\rm peak}$, which can be useful for direct comparisons with observations.

The peak time is roughly the time when the diffusion time scale of the ejecta $t_{\rm ej}^{\rm d}$ reduces to the dynamical time scale $t_{\rm dyn}$. In the optically thick case, the diffusion time scale is
\begin{equation}
    t_{\rm ej}^{\rm d}\sim \frac{R_{\rm ej}\tau_{\rm ej}}{c}=\frac{m_{\rm ej}\kappa}{4\pi R_{\rm ej}c}.
\end{equation}
We can roughly estimate the radius by $R_{\rm ej}\sim \beta_{\rm ej}ct$. Matching the diffusion time scale and the dynamical time scale, we have
\begin{equation}\label{eq:diffusion_ori}
    t_{\rm peak}=\sqrt{\frac{m_{\rm ej}\kappa}{4\pi c^2 \beta_{\rm ej}}}.
\end{equation}

The value of $\beta_{\rm ej}$ depends on whether the ejecta's kinetic energy around the peak time is dominated by the initial kinetic energy or the injected energy. Based on the two scenarios, we have the following derivations:

i) Initial energy dominated. In this case the velocity maintains its initial value $\beta_{\rm ej}=\beta_{\rm ej,0}$, so we can simply use eq. \ref{eq:diffusion_ori} with the proper scaling
\begin{equation}
    t_{\rm peak} = 4.87\left(\frac{m_{\rm ej}}{0.1M_{\odot}}\right)^{1/2}\left(\frac{\kappa_{\rm ej}}{1g\ cm^{-2}}\right)^{1/2}\left(\frac{\beta_{\rm ej,0}}{0.1}\right)^{-1/2}\ \rm d.
\end{equation}

ii) Injected energy dominated. In this case most of the injected energy (i.e., $\eta E_{\rm ini}$) is transformed to the kinetic energy of the ejecta, because the system remains optically thick before the peak time. We can calculate the velocity by $\beta_{\rm ej}=\sqrt{2\eta E_{\rm ini}/m_{\rm ej}c^2}$. Insert it into eq. \ref{eq:diffusion_ori}, we have
\begin{align}
    t_{\rm peak}=1.5&\left(\frac{m_{\rm ej}}{0.1M_{\odot}}\right)^{3/4}\left(\frac{\kappa_{\rm ej}}{1g\ cm^{-2}}\right)^{1/2}\left(\frac{E_{\rm ini}}{10^{53}erg}\right)^{-1/4}\nonumber\\
    &\times \eta^{-1/4}
    \ \rm d.
\end{align}

To determine which case is relevant, we can first try case 1, get $t_{\rm peak}^1$ and perform a consistency check, i.e., we can calculate the total injected energy up to this time
\begin{equation}
    E_{inj} = \frac{E_{\rm ini}t_{\rm peak}^1}{t_{\rm sd}+t_{\rm peak}^1},
\end{equation}
and then calculate the corresponding velocity. If it's smaller than $\beta_{ej, 0}$, the result is valid, otherwise we can move to the second case. 

From the above estimation we find the peak time is mostly dominated by ejecta mass and opacity, while other parameters have moderate impacts. This is in agreement with our anticipation, since these are the dominating parameters that determine when the ejecta becomes transparent.

\begin{figure}
    \centering
    \includegraphics[width=\columnwidth]{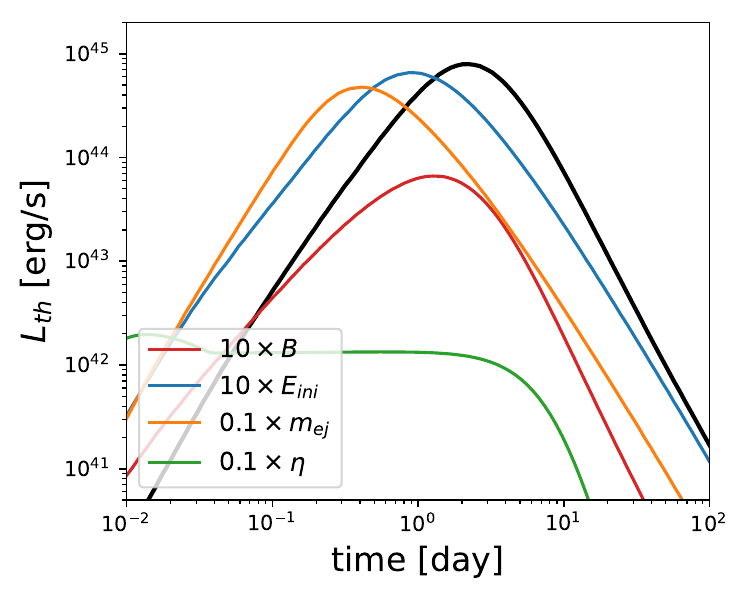}
    \caption{The comparison of optical light curves given different parameter combinations. The black line is the same as in previous figure, which serves as a ``fiducial" case. In the colored lines we change model parameters individually as indicated in the label corner. }
    \label{fig:compare_lth}
\end{figure} 

The numerical optical light curves for different combinations of parameters are shown in Fig. \ref{fig:compare_lth}, where we have compared the impact of different parameters by varying them against the previous ``typical" case. We find the above peak time estimation agrees well with our numerical result within a deviation smaller than a factor of 2. Additionally, the fact that the above two cases provides similar estimation indicates that the peak time of the magnetar-boosted KN is relatively universal, typically occurring within a few days after the merger.

Besides the peak time, we can also roughly examine the impact of different parameters on the peak luminosity, as shown in fig. \ref{fig:compare_lth}. We first examine the light curves produced by a indefinitely stable magnetar, where only the parameters $B$, $E_{\rm ini}$, and $m_{\rm ej}$ vary. We find that a large magnetic field can significantly suppress the luminosity due to increased pair production in the PWN, leading to higher opacity. The ejecta mass, on the other hand, doesn't strongly affect the results. This finding is mostly consistent with MP14.

Surprisingly, the peak luminosity appears to be insensitive to the initial rotational energy of the magnetar, which is essentially the available energy budget of the system. We have explored various parameter ranges and consistently found this result in our model. One explanation is that a larger rotational energy also corresponds to a higher spindown power, which leads to increased pair production similar to the case with a large magnetic field. Consequently, the increased energy budget is counterbalanced by the corresponding pair suppression. This can be understood by revisiting equation \ref{eq:initial_spindown} and equation \ref{eq:pair optical depth}, where the $E_{\rm ini}$ boosts the spindown power which increases the optical depth due to pair suppression. Another possible reason is that the spindown power is no longer dependent on $E_{\rm ini}$ when $t\gg t_{\rm sd}$. This can be seen by taking this limit for eq. \ref{eq:spindown}, \ref{eq:initial_spindown}, and \ref{eq:spindown_time}. This finding suggests that for indefinitely stable magnetars as merger remnants, the luminosity of the boosted KN is primarily determined by the magnetic field, regardless of the initial rotation of the magnetar.

In contrast to other parameters, $\eta$ has a much bigger impact on the light curve. Except for the early small bump caused by the collapse of magnetar, the rest of the light curve evolves like that of an adiabaticaly expanding shell, and fades away when it becomes optically thin. The peak luminosity drops by 3 orders of magnitude even if $\eta$ drops by just 1 order of magnitude, corresponding to $\eta E_{\rm ini}$ amount of total energy input.  The effect of $\eta$ is not same as trivially reducing $E_{\rm ini}$ accordingly, because the termination of central engine completely changes the acceleration process of the ejecta. 

\subsection{X-ray Spectrum evolution}
Another interesting behavior of this typical case is the frequency dependent evolution of the X-ray opacity, as presented in the lower panel of Figure \ref{fig:default}. It is important to note that the breakout time of X-rays is not a monotonically increasing function with photon energy up to 1 MeV. Instead, we find that photons with energies around 10 keV are the last to pass through the ejecta (see dotted pink line in the lower panel of Fig. \ref{fig:default}). This behavior is attributed to the nature of the ionization cross section as a function of photon energy. As we have explained in \S\ref{subsubsec:ionization}, the cross section extends from the ionization threshold energy to 1 MeV, which approximately follows a decaying power-law formula. Ions at low excited states generally have lower threshold energies, so the low energy photons can only ionize ions of low excited states. After the ejecta is highly ionized, only rare ions remain in low excited states. Consequently, the ejecta becomes nearly transparent to low-energy photons since they are unable to reach the threshold energy of the existing ions. On the other hand, high-energy X-rays can also easily pass through the ejecta due to their small cross section. It is the photons in the intermediate energy that experience the most absorption as they are capable of reaching the threshold energy while still having a relatively larger cross section. Therefore, our findings indicate a potential evolution of the X-ray spectrum. This behavior is not specific to iron-like elements but rather a common characteristic of heavy elements, as they generally follow a similar rule for cross sections. While our results may offer interesting observational predictions in the x-ray band, this is not the primary focus of our work, and we leave it for a future study. 

\subsection{The limit of forming an ultra-relativistic blast wave}
\label{sec:blast-wave}
As mentioned above, we find the system is generally unstable to the Rayleigh-Taylor instability. Though our estimation based on the linear growth rate may not be precise enough, it's still worthwhile to study the extreme limit in which the PWN matter completely leaks out from the ejecta. 

In this case, we assume the leaked energy forms an ultra-relativistic blastwave propagating into the surrounding environment. The blastwave accelerates electrons which produce synchrotron radiation, just like the case of a GRB. However, unlike regular GRBs, once the PWN leaks from the ejecta shell, there is no mechanism to confine the material into a narrow cone. As a result, the dynamics of the blastwave will be quasi-isotropic rather than jet-like. The light curve should be very similar to a GRB afterglow, except that it can be observed from all directions and exhibits no jet-break. We consider a simple analytic model of the blastwave (see, e.g. \citealt{2015PhR...561....1K}). To simplify the calculation, we only consider the slow cooling case where the synchrotron injection frequency, $\nu_{\rm m}$, is less than than the cooling frequency $\nu_{\rm c}$ and where the observed frequency falls within the range between the self-absorption frequency $\nu_{\rm a}$ and the cooling frequency $\nu_{\rm c}$. This frequency range is generally enough to encompass the optical waveband. The observed flux is
\begin{equation}
    f_{\nu} = 
    \begin{cases}
        f_{\rm\nu, max}(\nu/\nu_{\rm m})^{1/3} & \nu_{\rm a} < \nu < \nu_{\rm m} \\ 
        f_{\rm\nu, max}(\nu/\nu_{\rm m})^{-(p-1)/2} & \nu_{\rm m} < \nu < \nu_{\rm c},
    \end{cases}
\end{equation}
where $p$ is the electron energy distribution power-law index. The synchrotron peak frequency $\nu_{\rm m}$ is calculated by
\begin{align}
    \nu_{\rm m} &= 3.3\times 10^{12}(1+z)^{1/2}\ \mathrm{Hz} \\
    &\times \left(\frac{p-2}{p-1}\right)^2\left(\frac{\epsilon_{\rm B}}{0.01}\right)^{1/2}\left(\frac{\epsilon_{\rm e}}{0.1}\right)^2\left(\frac{E_{\rm iso}}{10^{52}\mathrm{erg}}\right)^{1/2}\left(\frac{t_{\rm obs}}{1\mathrm{d}}\right)^{-3/2},
\end{align}
where $\epsilon_{\rm e}$ and $\epsilon_{\rm B}$ are the fractions of internal energy of the blastwave converted to non-thermal electrons and magnetic fields, respectively. The peak flux is
\begin{align}
    f_{\rm\nu, max} &= 1.6\times 10^2(1+z)\ \mathrm{\mu Jy} \\
    &\times \left(\frac{\epsilon_{\rm B}}{0.01}\right)^{1/2}\left(\frac{E_{\rm iso}}{10^{52}\mathrm{erg}}\right)\left(\frac{n_0}{0.01}\right)^{1/2}\left(\frac{D_{\rm L}}{10^{28}\mathrm{cm}}\right)^{-2},
\end{align}
where $n_0$ is the ambient number density. Typically, there should be hydrodynamic breaks other than the spectral break, such as transition from coasting to deceleration, the lateral expansion (if the blast-wave is sufficiently anisotropic), and the transition from relativistic motion to Newtonian velocities. However, the first one is much earlier than the time of interest (tested for initial Lorentz factor of the blastwave $\Gamma > 100$), while the later two happen when the flux has significantly dropped off, thus not impacting our result. For simplicity, we don't include them here.

The microphysical parameters of this afterglow-like light curve should be similar to those of short GRBs. The ambient density should be relatively low with $n_0=10^{-3}$ - $10^{-2}$ cm$^{-3}$. We consider a fixed $\epsilon_{\rm e}=0.1$ since it is observationally constrained to be narrowly distributed between different GRBs \citep{BvdH2017}. As for $\epsilon_{\rm B}$, which is observationally less well-determined and may vary more from burst to burst, we consider a range between $10^{-3}$ to $10^{-2}$. Moreover, to make a fair comparison, we assume the energy of the blast wave matches with our ``fiducial" case, i.e., we fix the isotropic energy to $3\times10^{52}$ erg. 

\begin{figure}
    \centering
    \includegraphics[width=\columnwidth]{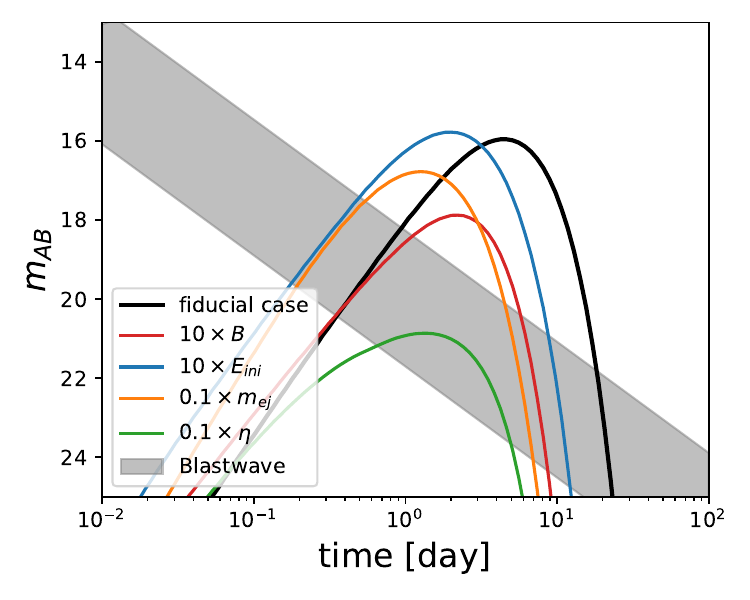}
    \caption{The flux of magnetar boosted KN. The parameters are the same as the previous figure. The redshift is set to 0.1, corresponding to a luminosity distance of $\sim$ 474 Mpc. The wave band is chosen to the $r$ band ($\sim$ 634 nm). In addition, we also plot the light curve of a blastwave in a grey shaded region assuming it is afterglow-like. The parameters of the shown blastwave are: $E_{\rm iso}=3\times 10^{52}$ erg, $n=10^{-3}-10^{-2}$ cm$^{-3}$, $\epsilon_{\rm e}=0.1$ and $\epsilon_{\rm B}=10^{-3}-10^{-2}$. The waveband is the same as for the KN.}
    \label{fig:compare_mAB}
\end{figure}

To compare the light curve in this scenario with the boosted-KNe model, we present them together in fig \ref{fig:compare_mAB}. The flux of the KN is calculated using eq. \ref{eq:flux}. The unit of flux is converted to AB magnitude for convenience. The KN parameters remain the same as those in Figure \ref{fig:compare_lth}, and the observable features of the flux match with the luminosity, so we don't repeat their description here. As we can see, the afterglow is in general comparable to or even brighter than the KN, though it peaks at a much earlier time. Moreover, since the blastwave is quasi-isotropic, the prospect of detection is not limited by the jet opening angle, thus significantly increasing the potential number of observable sources. If this scenario were correct, the detection rate of orphan afterglow would be expected to be very high, which is not evident in current observations (e.g. \citealt{2022ApJ...938...85H}). Therefore, the disruption of the KN ejecta because of the Rayleigh-Taylor instability is not a plausible explanation for the absence of detections of the transients following neutron star mergers. A quantitative simulation to explore the observed rate of the blast wave signatures is not pursued in this work.

\section{Constraint by absence of observation}\label{sec:observation}
In this section, we aim to compare our model with the current observations. As described in the previous section, the bright nature of magnetar-boosted KNe implies that they should be observable over a large cosmic volume. This suggests that the detection rate of these events could be comparable to, or even higher than, that of regular KNe, despite having a lower intrinsic event rate. The lack of confirmed detections puts strong constraints on either the parameters of the model or the characteristics of the remnants from neutron star mergers.

In this work, we perform Monte Carlo simulations to estimate the observational rates. However, certain model parameters, such as the initial rotational energy $E_{\rm ini}$ and energy extraction efficiency $\eta$, are not free to setup but depend on factors like the progenitor mass of the binary system and EoS. Furthermore, there is still a significant uncertainty in the EoS, which would introduce model dependency to the EoS based simulation. To address these considerations, we employ two approaches: a model-independent study and a model-dependent study. In the model-independent study, we generically evaluate the maximum detectable distance based on various parameter combinations, which serves as an indicator of the event rate. On the other hand, the model-dependent study begins with a population of binary systems, evolves them to remnants, and generates model parameters based on an assumed EoS. We then simulate observations using a sky survey strategy and collect the observed events to determine the detection rate. 

The model-independent study provides an overview of our model's predictions regardless of EoS, while the model-dependent study offers a quantitative result. The details of the two approaches are described below.

\subsection{Model independent study}

In this model-independent study we examine the observation potential of a parameter set indicated by its maximum detection distance. The maximum detection distance is roughly set by requiring the peak flux reaches a telescope's detection threshold. We assume the threshold is $m_{\rm AB}$=20.5, matching with the performance of ZTF \citep{2014SPIE.9147E..79S,2020PASP..132c8001D}. Similar to the previous section, the filter is set to r-band. We notice that a realistic confirmation of a detection requires more than one convincing data point, thus the peak flux should be slightly higher than detection threshold. We leave this effect to the model dependent study. 

The optical peak flux of the magnetar-boosted KN is primarily dependent on the values of magnetic field $B$ and energy extraction efficiency $\eta$, as explained in the previous section. In order to simplify our study, we fix the value of $m_{\rm ej}$ to $0.1M_{\odot}$ motivated by numerical simulations\citep{2019ApJ...880L..15M}. To illustrate the impact of both $B$ and $\eta$, we consider four cases: B=$10^{16}$ G, B=$10^{15}$ G, B=$10^{14}$ G, and B=$3\times10^{12}$ G. The first three cases account for the possible varying domain of magnetic field, while the last one serves as a self-consistency check, which will be described below. For each case, the maximum detection distance should be a function of $\eta$, which is solved by requiring the peak flux is equal to the detection threshold, i.e. $m_{\rm AB}$ = 20.5. To account for the minor influence of different $E_{\rm ini}$, each case is represented by a shaded region, encompassing $E_{\rm ini}$ values ranging from $10^{52}$ erg to $10^{53}$ erg. 

\begin{figure}
    \centering
    \includegraphics[width=\columnwidth]{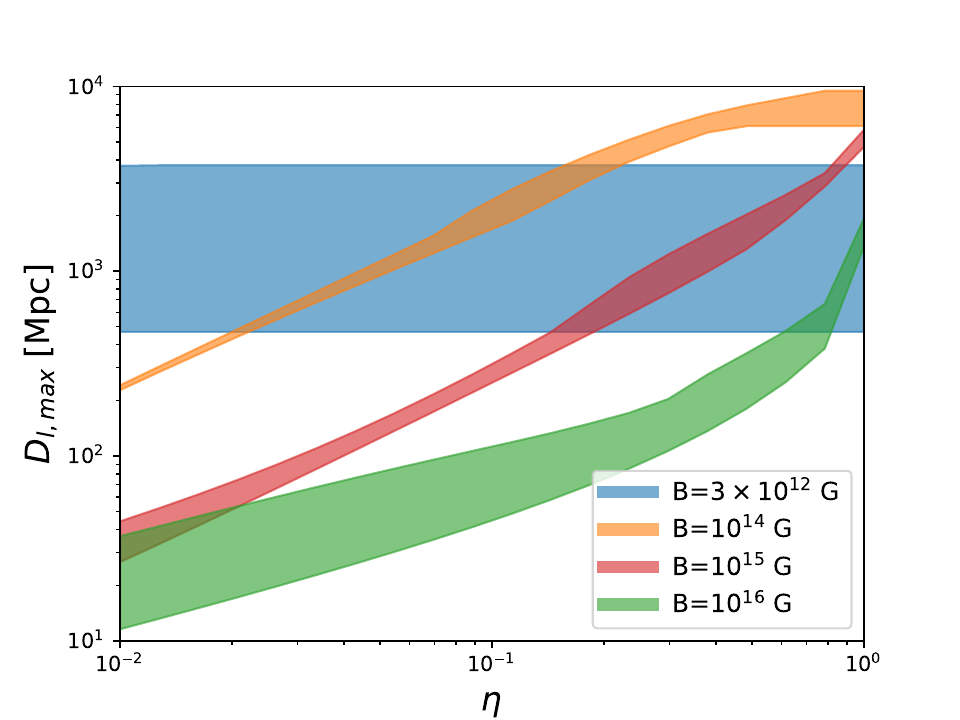}
    \caption{The maximum detectable distances of magnetar-boosted KN as a function of $\eta$ and with magnetar field strengths of $B=3\times 10^{12}$ G, $B=10^{14}$ G, $B=10^{15}$ G, and $B=10^{16}$ G. The shaded region is bracketed by the limits: $10^{52}\ \mathrm{erg}<E_{\rm ini}<10^{53}\ \mathrm{erg}$. The lowest $B$ case serves as a self-consistency check, where the peak luminosity should eventually reduce for such low magnetic fields, because the spin down timescale is much longer than the diffusion timescale. In this case the peak luminosity is also independent of $\eta$ since the magnetar never collapses in the time of interest.}
    \label{fig:model-independent}
\end{figure}

Our results are shown in Figure \ref{fig:model-independent}. As we can see, the peak luminosity of the KN increases with decreasing magnetic field strength, leading to the increase of maximum detectable distance. This is because reducing the magnetic field leads to a smaller spin down power, which produces less pairs in the PWN that suppress the energy injection. On the other hand, it is expected that the luminosity will eventually decrease as we keep decreasing the magnetic field, since the spin-down timescale could be significantly longer than the peak time, resulting in less energy injection while the ejecta is optically thick. To provide a self-consistency check, we verify this phenomenon by the B=$3\times 10^{12}$ G case. As shown in the blue shadow region, the maximum detectable distance indeed decreases as compared with the B=$10^{14}$ G case. 

The implications of this result will be discussed in \S \ref{subsec:constraints}. 

\subsection{Model dependent study}\label{subsec:MC}

In this model dependent study, we simulate the whole process from a given population of BNS systems to the detection rate of a boosted-KNe. Our method is explained in the following steps.

In the first step, we consider a population of BNS systems and let them merge and generate remnants. We assume that the population follows the Galactic neutron star mass distribution, which is approximated by a normal distribution with a mean value of $1.33M_{\odot}$ and a standard deviation of $0.09M_{\odot}$ \citep{2016arXiv160501665A,2016ARA&A..54..401O}. The properties of the remnants (which will be determined in the next step) depend on two parameters that are determined during this step: the baryonic mass and the initial rotational energy, i.e., $E_{\rm ini}$. The baryonic mass is obtained by summing the masses of the progenitor system and subtracting the ejecta mass. As previously mentioned, we anticipate the ejecta mass to be around $0.1 M_{\odot}$ if the merger remnant is a SMNS \citep{2019ApJ...880L..15M}. To account for potential uncertainties, we consider a uniform distribution ranging from 0.01 $M_{\odot}$ to 0.1 $M_{\odot}$, with an emphasis on the higher end of the range. The initial rotational energy is influenced by energy losses of the system during the differential rotation phase, which is not yet fully understood. We treat it as a free parameter scaled by the maximum allowed rotation, i.e., the Kepler rotation. Specifically, this free parameter is denoted as $E_{\rm ini}/E_{\rm kep}$.

In the second step, we calculate the fate of the merger remnant. A remnant can either collapse under its own gravity or survive for a period of time, depending on whether it reaches the threshold mass of uniform rotation. The survival of a remnant is highly sensitive to the equation of state (EoS), which can be roughly characterized by the Tolman-Oppenheimer-Volkoff (TOV) mass, denoted as $M_{\rm TOV}$. In our study, we consider two EoS: UU \citep{1988PhRvC..38.1010W} and SLY \citep{2001A&A...380..151D}, which correspond to $m_{\rm TOV}$ of $2.2M_{\odot}$ and $2.05M_{\odot}$, respectively. These values mostly cover the lower limit set by the most massive pulsars \citep{2013Sci...340..448A, 2021ApJ...915L..12F} and the upper limit constrained by the GW170817 (e.g., \citealt{2017ApJ...850L..19M} and \citealt{2018ApJ...858...74M}). To determine the evolution and status of a remnant based on its initial conditions, we employ the \texttt{RNS} code \citep{1995ApJ...444..306S}. Specifically, we calculate the threshold mass required for the formation of a SMNS, which depends on $E_{\rm ini}/E_{\rm kep}$, and compare it with the remnant mass calculated in the previous step to determine whether they can survive. We also calculate the critical rotational energy $E_{\rm crit}$ just before the collapse of the SMNS. This critical energy represents the end state of the SMNS. By comparing the initial energy $E_{\rm ini}$ with $E_{\rm crit}$, we can derive the energy extraction efficiency in our KN model $\eta = (E_{\rm ini} - E_{\rm crit})/E_{\rm ini}$. If the magnetar is indefinitely stable, we set this parameter to 1. Note that due to truncation errors in the \texttt{RNS} code, the obtained $m_{\rm TOV}$ values slightly deviate from the theoretical values. To avoid inaccuracies that could be caused by artificially scaling the results, we adopt the values provided by the code, which are approximately $2.17M_{\odot}$ for SLy and $2.07M_{\odot}$ for UU. This implementation doesn't change our conclusion. The precise value of $M_{\rm TOV}$ for UU implies a higher rate of SMNS formation than what we've considered, imposing a more stringent constraint on this EoS. On the other hand, the exact value of $M_{\rm TOV}$ for SLy suggests a shorter survival timescale for SMNS, which aligns with our conclusion.

In the third step, we generate a large set of event parameters. The ejecta mass $m_{\rm ej}$, initial rotational energy $E_{\rm ini}$, and energy extraction efficiency $\eta$ have been determined in the previous steps. To account for a range of magnetic fields, in this work, we assume a lower limit of $10^{14}$ G and upper limit of $10^{16}$ G with uniform distribution in logarithmic space. This assumption is based on the consideration of magnetic field amplification during the previous differential rotation phase of the magnetar, and the maximum magnetic field allowed for a stable magnetar configuration. The distribution of distances at which these events occur depends on the evolution of the BNS merger rate $R(z)$ with redshift. For simplicity, we assume that $R(z)$ is proportional to the sGRB rate and adopt the analytical approximation derived by \cite{2015MNRAS.448.3026W} (see equation 9 therein). The rate is then scaled to match the local BNS merger rate. Given other uncertainties in our model, we believe this assumption is sufficient. Using these parameter distributions, we generate a set of millions of parameter combinations. The total number of events is denoted by $N_{\rm tot}$.

Finally, we calculate the light curves based on our model. To address the uncertainty in ejecta opacity, we assume a uniform distribution of $\kappa$ in logarithmic space, ranging from 1 $\mathrm{cm^{-2}g^{-1}}$ to 10 $\mathrm{cm^{-2}g^{-1}}$, with an emphasis on the lower values. We then proceed to calculate the detection rate by selecting the light curves that can be detected. In order to compare with the ZTF observation, we designed a similar sky survey strategy. For each generated light curve, we select a series of time points in the r-band. The time interval between neighboring points is fixed at 3 days, in order to mimic the approximate cadence of the ZTF survey strategy (e.g., \citealt{2020ApJ...904..155A}). To introduce some variability, the time series is randomly shifted. A data point is considered "observed" if its flux exceeds the threshold value. Similar to the model-independent study, we assume the threshold flux to be $m_{\rm AB}$ = 20.5. We confirm the detection of an event if at least three data points are observed. We count all detected events and get the total number $N_{\rm det}$. The expected yearly detection rate is then calculated by
\begin{equation}
    R_{\rm det} = \frac{\Omega_{\rm fov}}{4\pi}\frac{N_{\rm det}}{N_{\rm tot}}\int_{0}^{r_{\rm max}}{\frac{r^2 R(z)}{1+z}\mathrm{d}r}
\end{equation}
where $r$ is comoving distance corresponding to redshift $z$ and $\Omega_{\rm fov}=47^{\circ}$ is the field of view of ZTF \citep{2020ApJ...904..155A}. In this simulation, the maximum distance $r_{\rm max}$ is set to be sufficiently large (corresponding to z$\sim$1.5) in order to cover the most optimal case (see fig. \ref{fig:model-independent}). The BNS merger rate is scaled to match the local rate $R(0)$ = 300 Gpc$^{-3}$ $yr^{-1}$, motivated by the recent constraint \citep{2022LRR....25....1M}. Different local BNS rates can easily be accommodated for by scaling the results given here.

\begin{figure}
    \centering
    \includegraphics[width=\columnwidth]{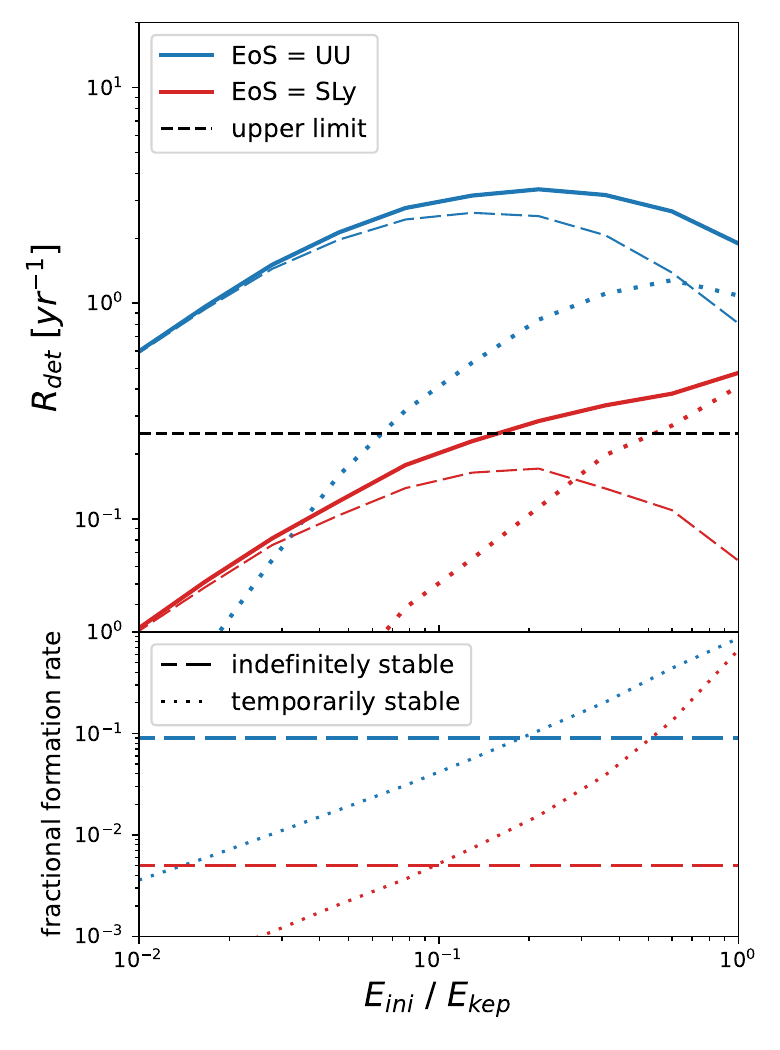}
    \caption{The simulated detection rate in the model dependent study. In the upper panel we show the detection rate (solid lines) as a function of the initial condition $E_{\rm ini}/E_{\rm kep}$ of SMNS assuming the equations of state of UU and SLy. The black dashed line is the approximate observational upper limit set by $1/T_{\rm ZTF}$. In the lower panel we show the fraction of BNS mergers leading to long-lived magnetar formation as a reference. To clarify the population of these events, we separate them to two classes. In both panels, the dashes lines are the events produced by indefinitely stable magnetars, and the the dotted lines are the temporarily stable magnetars. }
    \label{fig:model-dependent}
\end{figure}

The results are shown in fig. \ref{fig:model-dependent}. To demonstrate the population of detected events, the total detection rates (thick solid lines) are divided into two populations: events produced by indefinitely stable magnetars (thin dashed lines) and events produced by temporarily stable magnetars (thin dotted lines). The rates for each EoS are plotted in the upper panel\footnote{It is worth noting that the rate of boosted KNe caused by indefinitely stable magnetars decreases as the magnetar's rotational energy approaches the Kepler limit, as shown in the upper panel of fig. \ref{fig:model-dependent}. This result arises from the aforementioned effect: faster rotation leads to increased spindown power, which can potentially suppress the KN luminosity due to more pair production, rather than boosting it.}. In the lower panel, we show the formation rate of the two populations as a reference. 
Furthermore, considering the lack of confirmed detections, we can derive an approximate upper limit of $1/T_{\rm ZTF}$ yr$^{-1}$, where $T_{\rm ZTF}$ represents the effective operational time of the ZTF so far. Based on the reported effective operational time by the ZTF team in 2020 (approximately 2 years), we adopt the value of 4 years for this study. It is important to note that this upper limit is only an approximate estimation, and we will provide a more strict discussion below.

\subsection{Constraints on SMNS and the neutron star EoS}\label{subsec:constraints}

Our results place a strict constraint on the fate of SMNS. 

In Figure \ref{fig:model-independent}, our results reveal that if the magnetar is indefinitely stable, the optimal detectable distance can reach several to 10 Gpc. Even with strong magnetic field suppression (B=$10^{16}$ G), the maximum distance can still reach approximately 1 Gpc. Considering a neutron star merger rate of approximately 300 Gpc$^{-3}$ yr$^{-1}$, this implies that more than one thousand merger events can enter the detectable volume since the start of the optical survey. The absence of detections therefore suggests that the fraction of BNS mergers leading to SMNS formation must be smaller than $10^{-3}$, which is in contrast to the estimation based on the current constraints of EoS assuming $\eta=1$ (i.e., assuming Keplerian rotation at the birth time of SMNS).

A plausible explanation is that either most of merger remnants collapse during the differential rotation phase or, at the onset of uniform rotation, their energy is very close to the critical energy required to support them against collapse. The latter scenario corresponds to cases with $\eta \ll 1$ as shown in Figure \ref{fig:model-independent}.

This result can be more qualitatively seen in the model dependent study. In fig \ref{fig:model-dependent}. we can see that if the SMNS starts from Kepler rotation (i.e., $E_{\rm ini}/E_{\rm kep}=1$), both UU and SLy predicts detection number above the observational limit. For the EoS of UU, the expected detection rate is above the limit even with $E_{\rm ini}/E_{\rm kep}=10^ {-2}$. Therefore, our result tend to prefer SLy over UU. In other words, the $m_{\rm TOV}$ should be close to the lower limit of the current observational constraints. However, even with SLy, the expected detection rate is able to match with observation only when $E_{\rm ini}/E_{\rm kep}\lesssim 1$. Note if $E_{\rm ini}/E_{\rm kep}$ is too small, it will be impractical for the formation of SMNS, since their survival timescale would be negligible. If these events are the majority of the population, the events will be dominated by indefinitely stable neutron stars, instead of temporarily stable SMNS.

\begin{figure}
    \centering
    \includegraphics[width=\columnwidth]{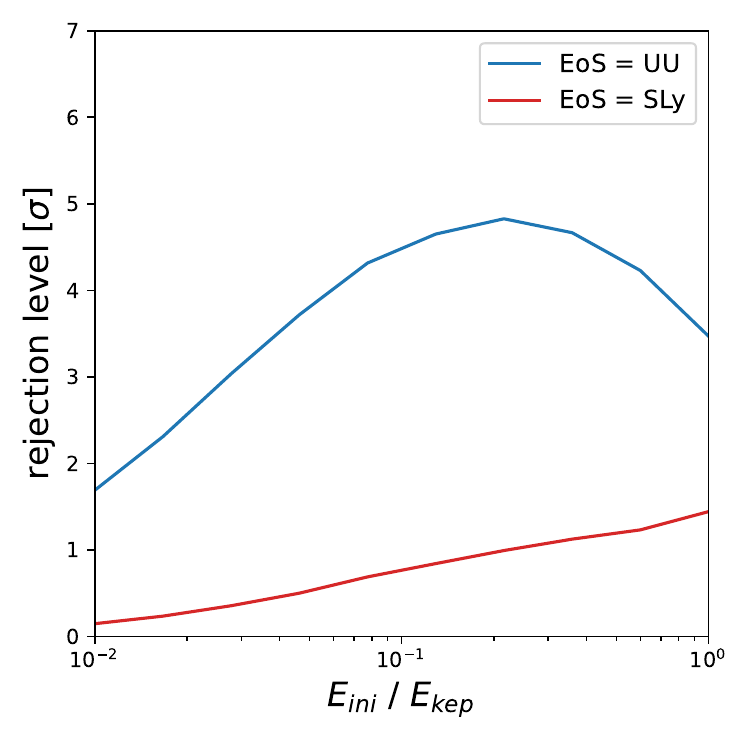}
    \caption{The rejection level of the given the initial rotation $E_{\rm ini}/E_{\rm kep}$ assuming equation of state of UU and SLy. The rejection probability is calculated assuming the number of detections follows a Poisson distribution. The probability is then converted to corresponding significance levels. Assuming EoS of UU, the scenario of SMNS is ruled out at a 3-$\sigma$ level. Assuming EoS of SLy, the initial rotation of SMNS at Kepler speed is ruled out at a 1-$\sigma$ level.}
    \label{fig:rejection-level}
\end{figure}

To provide a rigorous statistical analysis, we can perform a hypothesis test by assuming that the number of detections follows a Poisson distribution. The mean value for each $E_{\rm ini}/E_{\rm kep}$ case is the expected detection number over a period of $T_{\rm ZTF}=4$ years. The probability of observing 0 detection in a Poisson distribution is simply given by $e^{-\lambda}$, where $\lambda$ represents the mean value. We can then convert this probability into equivalent $\sigma$ levels in a normal distribution, which serves as an indicator of the "rejection level". The result is shown in fig. \ref{fig:rejection-level}.

In this result, we find that the scenario where the BNS merger remnants following the UU EoS forms a long-lasting SMNS is rejected at a significance level of $>3\sigma$ (see the blue line in fig. \ref{fig:rejection-level}) It is important to note that UU EoS itself is not ruled out, but the SMNS with high initial rotational energy is ruled out. 

On the other hand, for the SLy EoS, the long-lasting SMNS scenario is rejected at a significance level of $\gtrsim 1\sigma$ if the SMNS is initially rotating at the Keplerian speed. From these results, we can conclude that regardless of the EoS, when the merger remnant transitions into the uniformly rotating phase, it is unlikely to have a Keplerian rotation speed. This suggests that there must be some energy and angular momentum loss during the differential rotation phase.

This result is consistent with some recent studies and numerical simulations, where it is argued that the a merger remnant could collapse into a black hole before it becomes a SMNS, even if its mass allows to stabilize itself at Keplerian speed. This could happen if the remnant losses significant fraction of its angular momentum before it enters such phase \citep{2021ApJ...920..109B}. Alternatively, if the remnant rearranges its angular velocity profile during the differential rotation, such that the core slows down faster and initiates the collapse before reaching a uniform rotating configuration \citep{2022ApJ...939...51M}.

Our results are based on the assumption that the masses of merging neutron star binaries are similar to those observed in our Galaxy. Our constraint on EoS and initial rotation speed of SMNS can be relieved if the BNS mass distribution in the Universe is heavier than the Galactic distribution. However, this assumption leads to even more strict condition for SMNS formation. Therefore, the conclusion that SMNS are rare and short-lived objects is unchanged.

\section{Implications}\label{sec:implications}

\subsection{Implications on potential boosted kilonova candidates}
Besides the non-detection of boosted KN from sky surveys, there are some candidates (regular KNe) found in sGRB afterglow, such as GRB 130603B \citep{2013ApJ...774L..23B,2013Natur.500..547T}, GRB 060614 \citep{2015NatCo...6.7323Y,2015ApJ...811L..22J}, GRB 050709 \citep{2016NatCo...712898J} and GRB 080503 \citep{2009ApJ...696.1871P}. There is no clear evidence suggesting that the optical excess of these events require an additional energy source in the form of long-lived magnetars. In some of the events (e.g., GRB 130603B and AT2017gfo) there are strong limitations on the presence of long-lived magnetars, since their associated kilonovae show no signs of boosting. 

\begin{figure}
    \centering
    \includegraphics[width=\columnwidth]{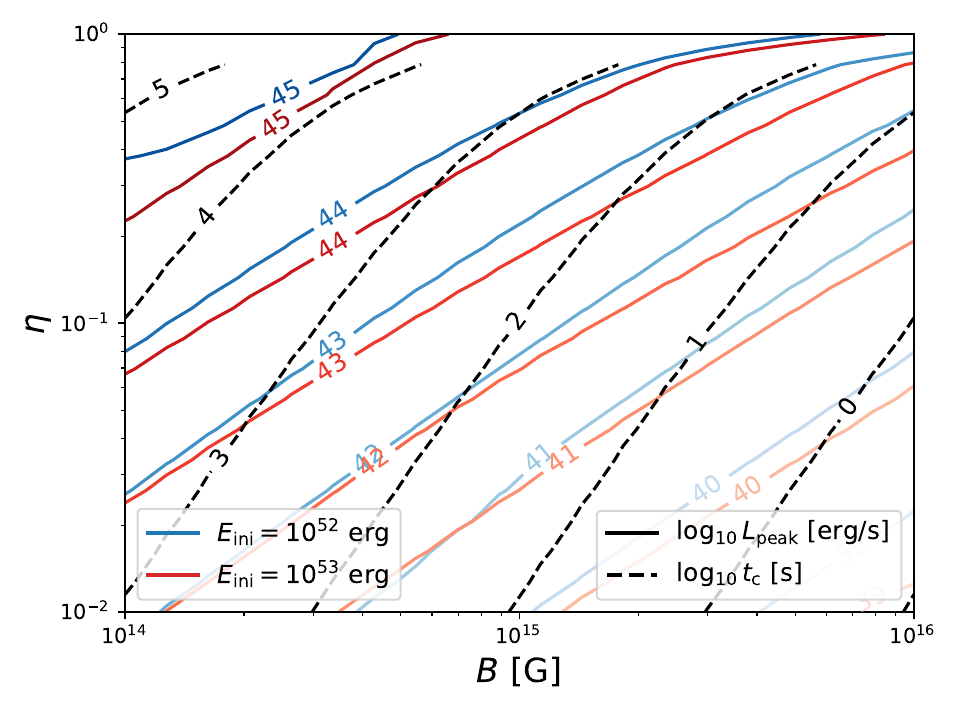}
    \caption{The peak luminosity and survival time of magnetar boosted KN with respect to two dominating parameters: the magnetic field $B$ and the energy injection efficiency $\eta$. The solid lines represents the KN peak luminosity for two values of the initial energy: $E_{\rm ini}=10^{52}$ erg and $E_{\rm ini}=10^{53}$ erg. The dashed lines represents the survival time $t_{\rm c}$ assuming $E_{\rm ini}=10^{53}$ erg. $t_{\rm c}$ will be 10 times larger for $E_{\rm ini}=10^{52}$. The ejecta mass and opacity are taken to be the typical values for a boosted KN: $m_{\rm ej}=0.1\ M_{\odot}$ and $\kappa=1$.}
    \label{fig:lpeak}
\end{figure}

To better demonstrate how the candidate events are in tension with the boosted KN model, we present a contour plot in Figure \ref{fig:lpeak}, showing the peak luminosity of boosted KNe as predicted by our model. This plot shows the variation of the peak luminosity with respect to the two dominant parameters, $B$ and $\eta$. Additionally, we also calculate the corresponding survival timescale $t_{\rm c}$ and present it on the same figure. We find that a long-lived magnetar that survives longer than the spindown timescale ($\eta\gtrsim 0.5$) and lasts for $\gtrsim 1000$ seconds generally boosts the KN to a luminosity $L_{\rm peak}> 10^{43}$ erg/s. Such a luminosity exceeds any confirmed or candidate KN known to date. In other words, to explain the lower luminosity of these events compared to our model predictions, one must assume a short survival timescale of the central magnetars, or rule out the magnetar explanation.

For instance, consider the only confirmed KN, AT2017gfo with a bolometric luminosity of $L_{\rm bol}\sim 10^{42}$ erg/s. If we assume a magnetar origin, in Figure \ref{fig:lpeak} we can find that even in the most extreme cases ($B\sim 10^{16}$ G), it still requires $\eta<0.5$, corresponding to a magnetar surviving less than the spindown timescale. In addition, the survival timescale in the allowed parameter region is less than an hour. This result further indicates that late time X-ray activity of GW170817 is unlikely to originate from a magnetar remnant. A similar conclusion can also be drawn regarding the other aforementioned candidate events in GRB afterglows.
Furthermore, the blast wave kinetic energy in those events {(e.g., $E_{\rm k}\lesssim 10^{50}$ erg for AT2017gfo, \citealt{2022ApJ...938...12B} and $E_{\rm k}\lesssim 10^{51}$ erg even for the beaming-corrected energy of the GRB afterglow blast wave in the same event, see \citealt{2021ARA&A..59..155M} and references therein)} is much lower than would be expected in case the nebula material had leaked out of the KN ejecta ($10^{52}$ erg - $10^{53}$ erg). 
Therefore, even if magnetars have ever existed in these events, they are unlikely to have been long-lived or responsible for late time activity.

The presence of a long-lived magnetar as the merger remnant of GW170817 is also disfavored by various other studies (e.g., \citealt{2017ApJ...850L..19M}; \citealt{2017ApJ...850L..24G}; \citealt{2018ApJ...854...60M}). Recently, a luminous candidate was found in GRB 200522A \citep{2021ApJ...906..127F}, which may serve as a candidate of boosted KN (see however \citealt{O'Connor2021} for a different interpretation). The luminosity ($\gtrsim 10^{42}$ erg/s) of this event, although brighter than the KN of GW170817, is still much fainter than the typical luminosity of a stable magnetar predicted in our model.

The small energy injection can also be explained if there are additional processes to slow down the magnetar except for dipole radiation. Possible candidates are gravitational waves and neutrino cooling. Previous works have argued that gravitational wave losses may dominate the spindown process if the ellipticity of the SMNS is sufficiently large \citep{2013PhRvD..88f7304F, 2013ApJ...779L..25F, 2018ApJ...861L..12L, 2018ApJ...860...57A}. Since these energy dissipation processes do not inject additional energy into the ejecta, modeling these processes is similar to using a shorter remnant survival timescale, i.e., a smaller $\eta$. This is because the small survival timescale compared with evolution timescale will hide the details of energy injection so that the result is insensitive to the spin down power index. However, in order to achieve the desired level ($\eta \ll 1$) as constrained by observations, these energy extraction processes must contribute approximately 10 times more energy than the dipole power. It remains uncertain whether this can be accomplished through gravitational waves or neutrinos. In particular, if such amount of loss is due to gravitational wave radiation, it will lead to unstable magnetic field configuration in the magnetar. Even if it happens, such powerful energy lose also implies a short survival timescale of the SMNS. In fact, recent studies (e.g. \citealt{2022MNRAS.516.4949S}) that incorporate gravitational wave emission have reached conclusions consistent with our work.

\subsection{Implications on Orphan GRB afterglow}

The lack of observational evidence for powerful KNe does not necessarily imply the absence of a powerful energy injection by the SMNS. It is plausible that the PWN energy is not transferred into the ejecta. Instead, the nebular-ejecta interface may turn out to be violently unstable to the RT instability which produces holes through the ejecta where PWN matter can escape. As we have mentioned before, the escaped energy will form a blastwave which produces powerful emission similar to the GRB afterglow, which implies an overwhelming number of orphan afterglow in contrast to observations. However, as our simulation indicates, there is still a chance that some merger remnants form indefinitely stable magnetars with low initial rotation energy. If prone to the RT instability, they may serve as a potential population of orphan afterglows. 

\subsection{Implications on progenitors of Fast Radio Bursts}

Fast rotating magnetars as BNS merger remnants are also considered as possible sources of fast radio bursts (FRB).There are two important constraints on this scenario. One is that the dispersion measure (DM) from the source can't exceed the total DM in FRBs, which is typically a few hundred pc cm$^{-3}$. In our model, the PWN is rich of pairs and the ejecta is highly ionized, implicating a very large DM, which may not be compatible with observations. {The other is the free-free absorption of radio waves.}

We first calculate the DM following
\begin{equation}
    DM=\int_{r_0}^{R_{\rm ej}} n_{\rm e} \mathrm{d}r\approx n_{\rm e}(R_{\rm ej}-r_0),
\end{equation}
where $n_{\rm e}$ is the electron number density along the path and $r_0$ is the site where FRB is generated. Depending on models, the FRB is either produced near the magnetar or in the magnetar wind, so the limit of $r_0$ is $0 < r_0 < R_{\rm n}$. However, in our model, the PWN is optically thick before the light curve peak. This means that any FRB produced near the magnetar can't escape from the PWN before the diffusion timescale (around peak time) due to Thompson scattering. Therefore, we assume $r_0=R_{\rm n}$, implying that FRBs are produced at the outer layer of PWN, and the source of DM is purely from the ejecta. In the ejecta, the number density of free electrons is $n_{\rm e} = 26 f_{\rm e}\rho_{\rm ej}/(56m_{\rm p})$ which are produced by the photoionization. Note that if an FRB is produced near the magnetar after the light curve peak, it should likely be able to escape. However, since the pair density calculated by eq. \ref{eq:pair density} is overestimated in this stage (though it doesn't affect the luminosity calculation), it can't be used to calculate the DM of the PWN. Our estimate of the DM (i.e., considering the ejecta only) at this stage should be regarded as a {\it lower limit}.

The free-free absorption optical depth of the radio waves places another constraint. For a similar reason as above, we also only calculate the optical depth in the ejecta $\tau_{\rm ff,\nu}^{\rm ej}=\alpha_{\rm ff, \nu}\Delta_{\rm sh}$, where the absorption coefficient $\alpha_{\rm ff, \nu}$ (see \citealt{2013LNP...873.....G}) is the sum of the contributions from all ion species 
\begin{equation}
    \alpha_{\rm ff, \nu} = 0.018\frac{n_{\rm e}\sum_{i=1}^{i=27}(i-1)^2n^{\rm i}}{T_{\rm ej}^{3/2}\nu^2}\Bar{g}_{\rm ff}.
\end{equation}
One may refer to Table \ref{tab:rad_symbols} for the definition of the symbols. Also note that $i=1$ corresponds to neutral atoms in our definition, meaning that ions at $i^{\rm th}$ ionization state have charges of $i-1$. In this work we assume the Gaunt factor is $\Bar{g}_{\rm ff}=1$. The frequency is set to $\nu$=1 GHz. The radio waves are unable to escape if the ejecta is optically thick to free-free absorption.

\begin{figure}
    \centering
    \includegraphics[width=\columnwidth]{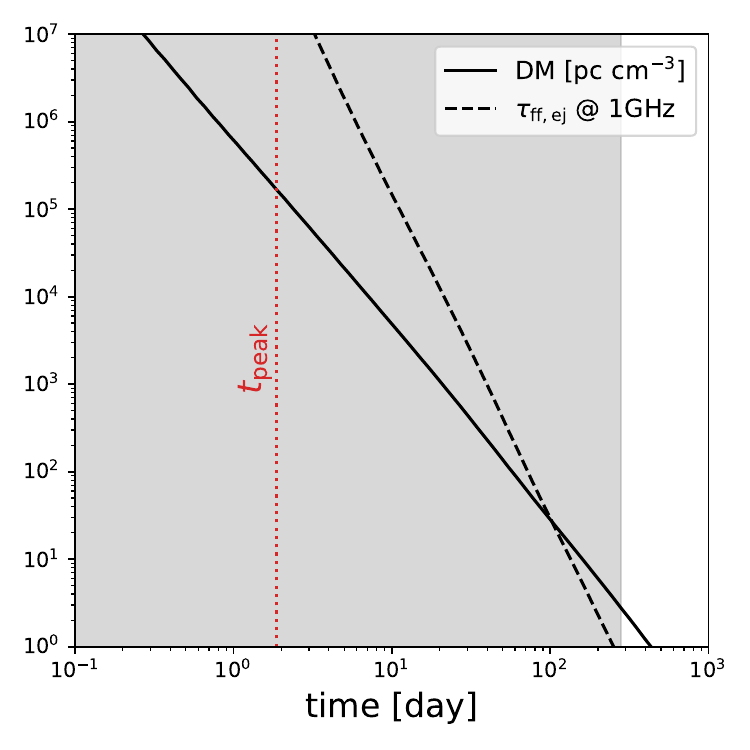}
    \caption{The dispersion measure (shown in the solid black line) and free-free optical depth (calculated at 1 GHz, shown in the dashed black line) of the ejecta shell along the evolution. The red dotted line is the peak time of the light curve for reference. The model parameters are taken to be same as fig. \ref{fig:default}. Note the typical DM observed in FRBs, which may be attributed to its local environment, is less than a few hundred. Since the estimated DMs are much greater than observed values for FRBs, and the ejecta is optically thick to radio waves, our result indicates that a FRB is not likely to form in the first year after the merger. The grey shaded region indicates the time period ruled out for FRB production owing to the free-free optical depth.}
    \label{fig:dm}
\end{figure}

Our results are shown in fig.\ref{fig:dm}. The parameters of the model are taken to be the same as in fig. \ref{fig:default}. We can clearly see that even if we only consider the DM from ejecta, it reduces to values consistent with observations after at least 10 to 100 days. Moreover, the situation becomes even more constraining when considering the free-free absorption optical depth, since the ejecta gets transparent only after $\sim$ a year (note that at this time the DM from the ejecta is $\lesssim 4\mbox{pc cm}^{-3}$ and as such is no longer constraining). These results indicate that FRBs can't escape from the ejecta for at least a year after the merger, thus challenging the merger model.

We caution that if the ejecta is disrupted due to the RT instability, the above calculation of DM no longer holds, since the pair cascade won't be triggered. In this case, we should also see a non-thermal multi-waveband radiation from the blastwave, and FRB radiation (if produced) may also escape more easily. We also caution that these arguments do not apply to the possibility of an FRB forming prior to the merger of BNS, e.g., a FRB produced by the interaction between the magnetospheres of the two neutron stars \citep{2021MNRAS.501.3184S, 2022MNRAS.515.2710M}.

\subsection{Implications on future Multi-messenger observation}
Magnetar-boosted KNe also serve as electromagnetic counterparts of gravitational waves. Our result (i.e., fig \ref{fig:model-independent}) has shown that their bright nature allows for a maximum detection distance of a few Gpc in the most optimal cases. This distance is much larger than the horizon distance of BNS mergers for LIGO's GW detectors in the current and future planed observing run. Our study indicates that such events are very rare, so we generally do not expect a detection as a counterpart of the gravitational waves. That being said, if such events indeed happen, considering their immense brightness, they are very likely to be detected. Such a detection would be extremely useful for constraining SMNS formation.

The detection of a booted KN will place a very strict constraint on the magnetar model. The fate of merger remnants can be sensitively constrained by the peak luminosity, which mostly depends on the survival timescale of the SMNS. The type of central engine (i.e., a stable magnetar, a SMNS, or a black hole) can be constrained by the slope of the light curve, which asymptotically follows the power of energy injection. 

As mentioned above, a long-lived SMNS is also a source of gravitational waves if it has some ellipticity. If its spin is fast and its survival timescale is long enough, the collected cycles of waves may be enough for detection before it collapses. The detection prospect sensitively depends on the ellipticity, which further depends on the type of instability during the rotation and the EoS (see \citealt{2015PASA...32...34L} for a review). Our result indicates that SMNS can't be long-lived, thus the gravitational waves from this stage are unlikely detectable. The detection of gravitational waves in this stage will imply a completely different interpretation of the lack of boosted-KN.

\section{Summary and Conclusion}\label{sec:summary}
In this study we built a magnetar-boosted KN model with a detailed description of relevant physical processes such as the collapse of the magnetar central engine and find its effects on the peak brightness of the light curve. To compare with observations, we conduct both a model dependent and model independent studies to estimate the detection rate. We find that the only way to match the non-detection result is to require an energy injection that is significantly smaller than the rotational energy of a maximally rotating remnant, which means that the SMNS (if formed) is likely to be short-lived in the vast majority of the cases. 

In principle, the system under consideration, a light PWN confined by the heavier KN ejecta is prone to the  Rayleigh-Taylor instability. Indeed the linear growth rate of a mode with wavelength comparable to the ejecta thickness is typically faster than the expansion timescale of the system. This leads to the possibility that the PWN breaks out from the KN ejecta driving a relativistic blast wave into the ambient gas. We find the blast wave produces an afterglow-like emission, which is quasi-isotropic and of comparable or brighter than the boosted KN signal. Therefore, the uncommon detection of orphan afterglows implies that this is not a common occurrence.

Our result has several implications. The statistically short-lived SMNS either means the initial rotation of a newly born SMNS is much slower than the Kepler rotation, or they collapse into a black hole before most of energy are released. One possible reason is that during the differential rotation phase, the angular momentum transfer from the core to the outer shell is fast, such that the SMNS collapse into a black hole before it establishes uniform rotation. Other possibilities are additional energy extraction mechanisms, such as the gravitational wave radiation or neutrino cooling, or alternatively, a heavier neutron star mass distribution. These all lead to early collapse. Due to their short survival timescale, we do not expect to see the magnetar-boosted KNe in LIGO's upcoming observation run.

\section*{Acknowledgements}
We thank Brian D. Metzger for his helpful discussion and suggestions. We also thank the anonymous referee's useful comments. HW and DG acknowledge support from the NSF AST-2107802, AST-2107806 and AST-2308090 grants. PB was supported by a grant (no. 2020747) from the United States-Israel Binational Science Foundation (BSF).

\section*{Data Availability}

The data underlying this article will be shared on reasonable request to the corresponding author.



\bibliographystyle{mnras}
\bibliography{mnras_template} 





\bsp	
\label{lastpage}
\end{document}